\documentclass[aps,pra,twocolumn,superscriptaddress]{revtex4-2}

\usepackage{graphicx,amsmath,amssymb}
\usepackage{hyperref}

\newcommand{\C}{$^{\circ}\mathrm{C}$}
\newcommand{\unit}[1]{\mathord{\mathrm{#1}}}
\newcommand{\ket}[1]{\lvert #1 \rangle}

\begin{document}

\title{Zeeman slowing of a group-III atom}
\author{Xianquan Yu}
\affiliation{Centre for Quantum Technologies, National University of Singapore, 3 Science Drive 2, Singapore 117543}
\author{Jinchao Mo}
\affiliation{Department of Physics, National University of Singapore, 2 Science Drive 3, Singapore 117551}
\author{Tiangao Lu}
\affiliation{Centre for Quantum Technologies, National University of Singapore, 3 Science Drive 2, Singapore 117543}
\author{Ting You Tan}
\affiliation{Department of Physics, National University of Singapore, 2 Science Drive 3, Singapore 117551}
\author{Travis L. Nicholson}
\email{nicholson@nus.edu.sg}
\affiliation{Centre for Quantum Technologies, National University of Singapore, 3 Science Drive 2, Singapore 117543}
\affiliation{Department of Physics, National University of Singapore, 2 Science Drive 3, Singapore 117551}

\begin{abstract}
    We realize a Zeeman slower of an atom in main group III of the Periodic Table, otherwise known as the ``triel elements.'' Despite the fact that our atom of choice (namely indium) does not have a ground state cycling transition suitable for laser cooling, slowing is achieved by driving the transition $\lvert 5P_{3/2},F=6 \rangle \rightarrow \lvert 5D_{5/2},F=7 \rangle$, where the lower-energy state is metastable. Using a slower based on permanent magnets in a transverse-field configuration, we observe a bright slowed atomic beam at our design goal velocity of 70 m/s. The techniques presented here can straightforwardly extend to other triel atoms such as thallium, aluminum, and gallium. Furthermore, this work opens the possibility of cooling group-III atoms to ultracold temperatures.
\end{abstract}

\date{\today}
\maketitle

\section{Introduction}
Since the achievement of Bose-Einstein condensation of laser cooled atoms in 1995, nearly all quantum degenerate gas experiments have been based on alkalis, alkaline earths, or lanthanides \cite{Schr2021}. Meanwhile, most of the Periodic Table remains unexplored in the quantum degenerate regime. One such unexplored class of atoms are the triels (Periodic Table main group III), which contains the atoms B, Al, Ga, In, and Tl. Unlike the $S$-orbital ground states of alkalis and alkaline earths, or the high angular momentum ($L = 5,6$) ground states of erbium and dysprosium, the anisotropic $P$-orbital ground states of group-III atoms distinguish themselves as intermediate cases. Although no group-III atom has been cooled to ultracold temperatures, quantum gases of these particles would have many interesting properties. Like alkaline earths, In and Tl have narrow-linewidth electronic transitions at wavelengths amenable to stable laser technology; however, unlike alkaline earths, triels also have ground state magnetic Feshbach resonances. Therefore, triel atoms could be probed with atomic clock resolution while offering the many-body control of alkali atoms.

Studies of optical forces on triels have largely focused on their application to nanofabrication \cite{McGowan95,Rehse2004,Kloter2008}. Optical forces have been observed in Al \cite{McGowan95}, Ga \cite{Rehse2004}, and In \cite{Kloter2008,Kim2009}, whereas Tl has been suggested as a laser cooling candidate \cite{Fan2011}. However, standard cooling techniques for ultracold gas production, such as Zeeman slowers and magneto-optical traps, have never been demonstrated in triel atoms. One drawback to triels (compared to alkalis and alkaline earths) is the lack of cycling transitions suitable for laser cooling in their $P_{1/2}$ ground states. However, the $P_{3/2}$ first excited states are long lived and the $P_{3/2} \rightarrow D_{5/2}$ line is a closed cycling transition that is generally amenable to laser cooling \cite{McGowan95,Rehse2004,Kim2009}.

To produce a slow beam of group-III atoms suitable for laser cooling to ultracold temperatures, techniques such as Zeeman slowing \cite{Phil1982} or 2D magneto-optical trapping \cite{dieckmann1998,lamporesi2013} would need to be realized for these particles. In this paper, we report a Zeeman slower of a triel atom, namely indium. Our approach laser cools indium atoms on the $\ket{5P_{3/2},F=6} \rightarrow \ket{5D_{5/2},F=7}$ transition. We use a permanent magnet Zeeman slower in a transverse configuration \cite{Ovch2007}, and the slower's magnet array is designed with a simulation predicting a 70 m/s final beam velocity. Using five laser wavelengths for state preparation and repumping, we observe a slowed indium beam with a velocity at our design goal. This work paves the way for realizing quantum degenerate gases of indium and other group-III atoms.

\section{Indium beam slowing scheme and apparatus}

\begin{figure}[htbp]
	\centering
	\includegraphics[width=\linewidth]{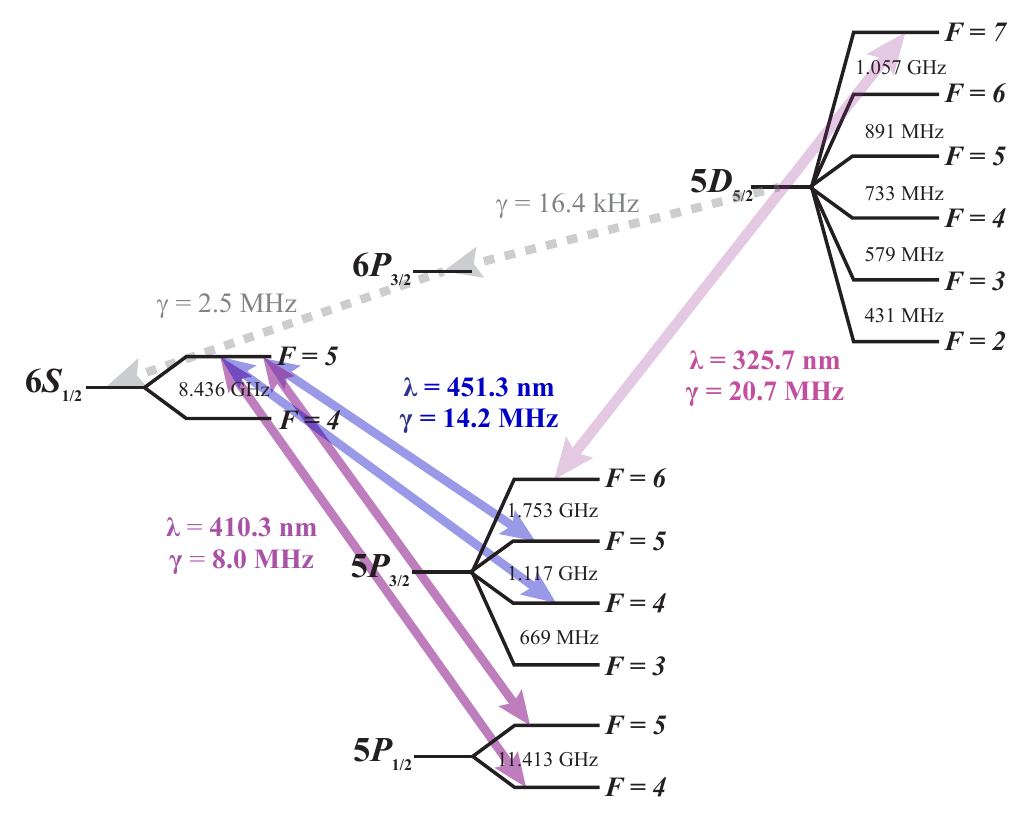}
	\caption{Energy levels of $^{115}\mathrm{In}$ \cite{Eck1957,Gunawardena2009,Kim2009,Zimmermann1970,Safr2007}. Transition wavelengths are denoted with a $\lambda$ and the corresponding linewidths are denoted with a $\gamma$. Our cooling scheme is based on the 326~nm $5P_{3/2} \rightarrow 5D_{5/2}$ transition, which has a $\gamma = 20.7 \, \mathrm{MHz}$ linewidth \cite{Kim2009}. Atoms can be pumped into the $\ket{5P_{3/2},F=6}$ cooling state with two lasers at 410~nm and another two at 451~nm. The lifetime of the $5P_{3/2}$ metastable state is predicted to be 10 s \cite{Sahoo2011}.}
	\label{fig:In_levels}
\end{figure}

Indium has two stable isotopes, $^{113}\mathrm{In}$ (4.3\%) and the more abundant $^{115}\mathrm{In}$ (95.7\%), the latter of which we work with. Both are bosons and have nuclear spin of $9/2$. The dipole allowed $5P_{1/2} \rightarrow 6S_{1/2}$ transition out of the indium ground state can be addressed with diode lasers, but it is not suitable for laser cooling because $6S_{1/2}$ decays rapidly to $5P_{3/2}$ (Fig. \ref{fig:In_levels}). Despite this, beams of ground state indium atoms were previously transversely cooled by driving both the $5P_{1/2} \rightarrow 6S_{1/2}$ and $5P_{3/2} \rightarrow 6S_{1/2}$ transitions \cite{Kloter2008}. Unfortunately, the energy level complexity and the resulting dark states rendered this attempt inefficient. A later attempt to transversely cool an indium beam using the $5P_{3/2} \rightarrow 5D_{5/2}$ transition yielded better results \cite{Kim2009}. Cooling on this transition was also demonstrated for the group-III atoms Al \cite{McGowan95} and Ga \cite{Rehse2004}, but only for the small thermal $P_{3/2}$ population of hot effusive beams.

In this work, to cool on the $\ket{5P_{3/2},F=6} \rightarrow \ket{5D_{5/2},F=7}$ transition, we use the following scheme to prepare atoms in the lower cooling state (Fig. \ref{fig:In_levels}). First, atoms are driven into the $5P_{3/2}$ hyperfine manifold with a pair of 410.3~nm external-cavity diode lasers (ECDLs) addressing the $\ket{5P_{1/2},F=4,5} \rightarrow \ket{6S_{1/2},F=5}$ transitions. Atoms are then pumped into $\ket{5P_{3/2},F=6}$ with another pair of ECDLs at 451.3~nm. The 325.7~nm cooling laser is generated by a 1302.8~nm ECDL that seeds a Raman fiber amplifier. The amplifier output is frequency quadrupled to achieve hundreds of milliwatts of useful 325.7~nm laser power.

\begin{figure*}[htbp]
	\centering
	\includegraphics[width=\linewidth]{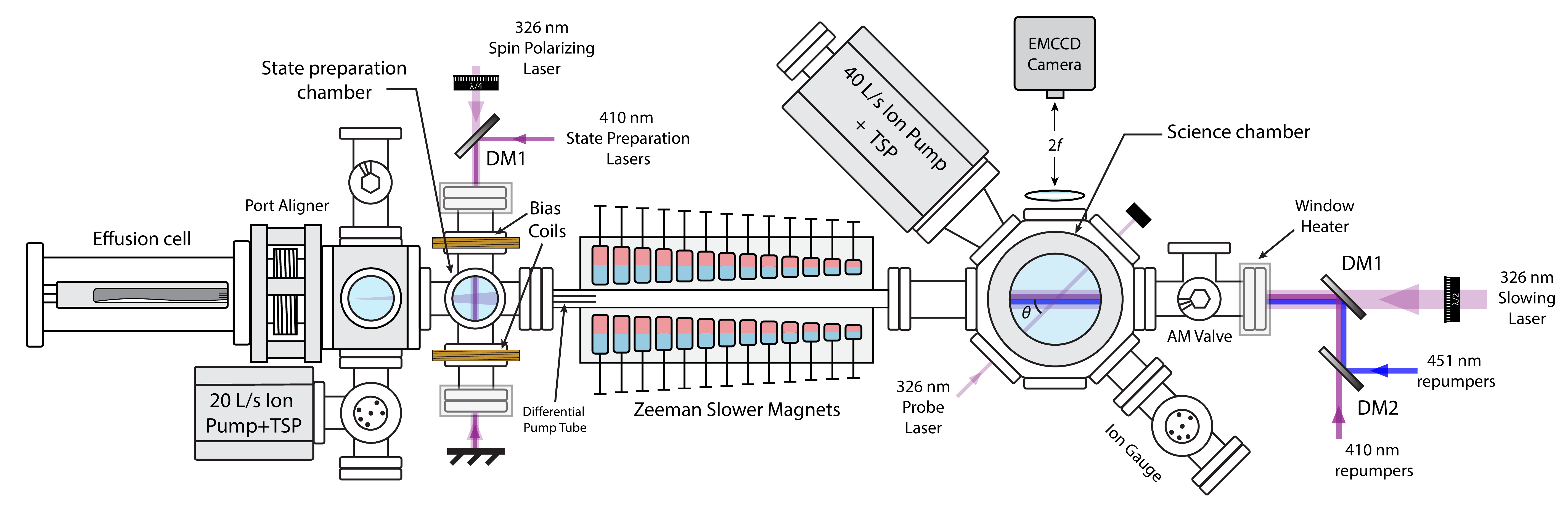}
	\caption{Schematic of the indium Zeeman slower apparatus. A hot indium atomic beam is produced from an effusion cell running at 800 \C. Indium atoms from the oven are pumped into the $\ket{5P_{3/2},F=6,m_F=6}$ cooling state in the state preparation chamber by two 410~nm lasers and a 326~nm spin polarization laser. The atomic beam is then decelerated by a transverse permanent magnet Zeeman slower. Slowing is accomplished with a 326~nm laser, two 410~nm repumpers, and two 451~nm repumpers, all of which are combined using dichroic mirrors (DM) and sent through the Zeeman slower. An independent 326~nm probe laser intersects with the atomic beam at $\theta = 45^\circ$ in the science chamber. To measure the longitudinal atomic velocity distribution, fluorescence is collected with an EMCCD camera.}
	\label{fig:apparatus}
\end{figure*}

Our system produces an indium beam using an effusion cell (Fig. \ref{fig:apparatus}). The cell contains 22 g of indium and operates at 800 \C, which results in an indium vapor pressure of $~10^{-3}\ \unit{Torr}$ inside the effusion cell crucible. The output of the crucible has a 3D printed titanium microchannel array that helps collimate the indium beam. The array is 5 mm in diameter, with 200 microchannel tubes that are each 1 cm long and 200 $\unit{\mu m}$ in diameter.  The microchannel array is heated to 900 \C\ to prevent clogging.

We estimate that less than 10\% of the atoms emerging from the effusion cell are in $5P_{3/2}$; therefore, the atoms pass through a state preparation chamber before they enter the Zeeman slower. The vacuum pressure in this chamber is at the mid $10^{-9}$ Torr level when the effusion cell is at full temperature. State preparation consists of three laser beams. Two of them are at 410~nm, which drive the $\ket{5P_{1/2},F=4,5} \rightarrow \ket{6S_{1/2},F=5}$ transitions. We have observed that atoms decay from $\ket{6S_{1/2},F=5}$ to the $\ket{5P_{3/2},F=6}$ cooling state with a 60\% branching ratio, which agrees with our estimation based on angular momentum factors. Additionally, we include a 326~nm circularly polarized laser and a pair of bias coils to spin polarize the atoms into $\ket{5P_{3/2},F=6,m_F=6}$, which improves the efficiency of our Zeeman slower \cite{Ali2017,Lison1999}.

After the state preparation chamber, atoms pass through a differential pump tube and then into a $470\ \unit{mm}$ long Zeeman slower vacuum tube with an inner diameter of $16\ \unit{mm}$. The Zeeman slower magnets are N35 neodymium stacks arranged in a transverse field configuration \cite{Ovch2007,Rein2012,Hill2014} (Sec. \ref{sec:slower_design}). The Zeeman slower laser beam at 326~nm is linearly polarized, as is required for the transverse configuration \cite{Ovch2007,Rein2012,Hill2014,Chei2013,Ali2017,Mele2004}, and it has a beam area of $8\ \unit{mm} \times 6\ \unit{mm}$. This beam is gently focused \cite{Hopk2016} such that it has an area of $6 \ \unit{mm} \times 4\ \unit{mm}$ at the entrance of the Zeeman slower. Due to the population decay pathway $5D_{5/2} \rightarrow 6P_{3/2} \rightarrow 6S_{1/2}$ as well as off-resonant driving of $\ket{5P_{3/2},F=6} \rightarrow \ket{5D_{5/2},F=6}$, atoms have a small chance of decaying into hyperfine states that are not addressed by the slowing laser. Therefore, two 410~nm repumpers driving the $\ket{5P_{1/2},F=4,5} \rightarrow \ket{6S_{1/2},F=5}$ transitions and two 451~nm repumpers driving the $\ket{5P_{3/2},F=4,5} \rightarrow \ket{6S_{1/2},F=5}$ transitions are coaligned with the slowing laser.

Atoms emerge from the slower and enter the science chamber, which is held at the low $10^{-10}\ \unit{Torr}$ level. Here we measure the indium beam velocity distribution, which is probed with an independent 326~nm laser. This probe laser is based on a 1303~nm ECDL-seeded tapered amplifier, the output of which is sent to a waveguide doubler and then to a home-built BBO doubling cavity. The maximum 326~nm output of the BBO cavity is $10\ \unit{mW}$. The probe laser intersects with the indium beam at a $45^\circ$ angle, ensuring that it samples the longitudinal velocity distribution. Probe fluorescence is collected with a $2f$ imaging system focused onto an EMCCD camera.

\section{Slower Design}
\label{sec:slower_design}
For decades, Zeeman slowers were based on tapered solenoids \cite{Phil1982,Mole1997,Court2003,Hopk2016}, but in recent years permanent magnet Zeeman slowers have become a popular choice \cite{Ovch2007, Chei2011, Rein2012, Hill2014, Qiang2015, Ali2017, Wodey2021}. The particular permanent magnet slower design we have chosen is a transverse field slower based on a planar array of magnets \cite{Ovch2007,Rein2012,Hill2014}. This design is simple, effective, easy to adjust, and minimizes stray fields. Here we provide an overview of our Zeeman slower design.

An atom moving through a laser beam and a magnetic field experiences the force \cite{Metcalf2001}

\begin{align}
    \vec{F} = \frac{\hbar \vec{k} \Gamma}{2} \frac{s}{1+s+(\frac{2\Delta}{\Gamma})^2} ,
    \label{eq:optical_force}\end{align}
where $\Gamma = 2\pi\gamma$ is the transition natural linewidth in rad/s, $\vec{k}$ is the laser wavevector, $s$ is the saturation parameter, and 

\begin{align}
    \Delta = \Delta_0 + \vec{k} \cdot \vec{v} - \mu_{\textit{eff}}B/\hbar.
\end{align}
Here $\Delta_0 = \omega_{laser} - \omega_{atom}$ is the laser detuning (i.e., the difference between the laser and atomic resonance frequencies),  $\mu_{\textit{eff}} = (g_e m_e - g_g m_g)\mu_{B}$ the transition magnetic moment, $g_e$ ($g_g$) the Land\'e g factor of the excited (ground) state, $m_e$ ($m_g$) the magnetic quantum number of the excited (ground) state, $\mu_B$ the Bohr magneton, $\vec{v}$ the atomic velocity, and $B$ the external magnetic field magnitude. The optical force [Eq. (\ref{eq:optical_force})] maximizes when the magnetic field is chosen such that $\Delta = 0$, resulting in the optimal slower field

\begin{equation}
    B = \frac{\hbar}{\mu_{\textit{eff}}}\left(\Delta_0 + \vec{k}\cdot\vec{v} \right).
    \label{eq:ideal_field_1}
\end{equation}

Atoms moving through this field will experience a constant deceleration given by
\begin{equation}
    \vec{a} = \frac{\hbar \vec{k} \Gamma}{2m}\frac{s}{1+s} = \vec{a}_{max}\eta,
\end{equation}
where $\eta = s/(1+s)$ is the so-called ``design parameter'' determined by the slowing laser intensity. The solution for the classical motion of the atoms when $\Delta = 0$ is 

\begin{equation}
    v = \sqrt{v_0^2 + \left(v_f^2 - v_0^2 \right)\frac{z}{L}},
    \label{eq:atomic_vel_profile}
\end{equation}
where $v_0$ is the initial velocity, $z$ is the position of the atoms along the Zeeman slower, $L = \frac{v_0^2-v_f^2}{2 \eta a_{max}}$ is the length of the slower, and $v_f$ is the desired final atomic velocity. Therefore, the optimal field is

\begin{equation}
    B = \Delta B \sqrt{1-\left(\frac{v_0^2-v_f^2}{v_0^2}\right)\frac{z}{L}}+B_0,
    \label{eq:ideal_B_field}
\end{equation}
where $\Delta B = \hbar k v_0/\mu_{\textit{eff}}$ is the full magnetic field span and $B_0 = \hbar \Delta_0/\mu_{\textit{eff}}$ is the magnetic field offset.

The parameters $\Delta B$, $\eta$ and $B_0$ are determined by experimental constraints. With the atomic species and transition selected, $\Delta B$ is entirely determined by $v_0$. It is desirable to make $v_0$ as large as possible since all atoms with velocities less than $v_0$ can be slowed; however, one must also try to prevent $L$ (which increases as $v_0^2$) from being too large since the transverse atomic velocity distribution causes particle loss that scales as $L^2$. We choose $v_0 = 450\ \unit{m/s}$, which addresses $\sim 37\%$ of the longitudinal Maxwell transition distribution \cite{ramsey1956,huebner2000} emerging from the 800 $^\circ$C  effusion cell. This $v_0$ results in $\Delta B = 987\ \unit{G}$, which is a reasonable value for neodymium magnets.

The design parameter is determined by the slowing laser intensity $I$. In the case of In, the 326~nm cooling transition has a saturation intensity $I_{sat} = 78.3\ \unit{mW/cm^2}$. This relatively high saturation intensity (compared to alkalis and alkaline earths) makes it difficult to operate in the regime where $s = I/I_{sat}$ is much greater than 1, particularly when we account for the fact that 50\% of the slowing laser power is associated with a polarization helicity that is not useful for slowing \footnote{A transverse field slower requires the slowing laser to be linearly polarized perpendicular to the magnetic field; therefore, only one helicity of the resulting $\sigma^{+}$ and $\sigma^{-}$ polarization superposition is useful for slowing \cite{Mele2004,Ovch2007,Hill2014}}. However, making $\eta$ small would again increase $L$ and result in particle loss. We choose $s = 2/3$ ($\eta = 0.4$) for a good balance between available power and reasonable slower length. Additionally, we choose $v_f = 70\ \unit{m/s}$ because this is a reasonable value for future studies like magneto-optical trapping. These numbers result in $L = 356 \ \unit{mm}$.

\begin{figure}[htbp]
	\centering
	\includegraphics[width=\linewidth]{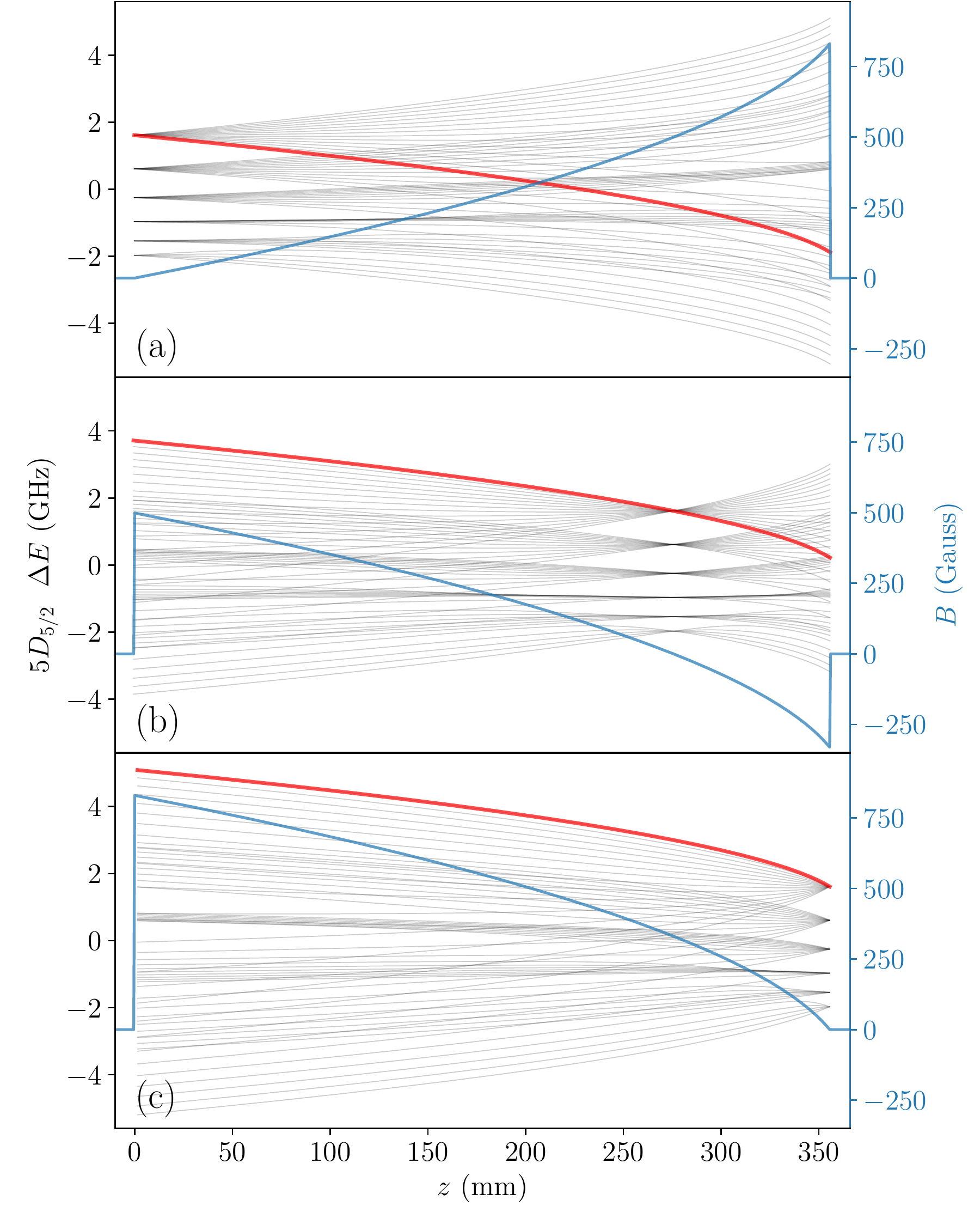}
	\caption{$5D_{5/2}$ energy states in different magnetic field configurations as a function of position along the Zeeman slower (black curves). Here we assume the ideal magnetic field of Eq. (\ref{eq:ideal_B_field}) with $v_f=70\ \unit{m/s}$. The red curves indicate the Zeeman shift that the $5D_{5/2}$ cooling state requires for slowing. The blue curves show the magnetic field profiles for the different Zeeman slower configurations. Panel (a) is the increasing field slower, (b) is the spin flip slower, and (c) is the decreasing field slower.}
	\label{fig:zeeman_spliting}
\end{figure}

To determine $B_0 = \hbar \Delta_0/\mu_{\textit{eff}}$, we consider three common Zeeman slower field configurations: the increasing field slower, the spin-flip slower, and the decreasing field slower \cite{Hopk2016}. These configurations work equally well for a two-level atom, but indium hyperfine structure complicates the story (Fig. \ref{fig:zeeman_spliting}). For the increasing field configuration \cite{Bagn1991,Joff1993,McCl2006,Marti2010,Chei2011,Ali2016} (also referred to as the ``$\sigma^-$ configuration''), indium atoms would need to be driven into the $\ket{5D_{5/2},F=7,m_F=-7}$ state; however, this state mixes with other $\ket{5D_{5/2},F=6,5,4,3}$ hyperfine levels in the required magnetic fields [Fig. \ref{fig:zeeman_spliting}(a)]. Driving these mixed states may cause atoms to decay to levels that are not resonant with the slowing laser and ruin the slowing efficiency \cite{gunter2004,Bell2010,Marti2010,Chei2011,Chei2013,Ali2016,Ali2017}, especially when the slowing laser contains both $\sigma^+$ and $\sigma^-$ components. Zeeman state mixing also occurs in the $\ket{5P_{3/2},F=6}$ cooling state when the magnetic field is between 350 G and 1300 G. Some alkali Zeeman slowers overcome this mixing by adding an offset to the magnetic field until all the Zeeman sublevels are fully split \cite{gunter2004,Marti2010,Chei2011,Ali2016}. With alkali atoms, only a modest magnetic field offset is needed; however with indium, an external field of 1600 to 2600 G would be required to fully split the Zeeman sublevels. According to our calculations, this high field cannot be achieved with neodymium magnets without considerable field inhomogeneity along the transverse atomic beam direction. 

Another popular Zeeman slower field configuration that results in much less state mixing at reasonable fields is the spin-flip configuration \cite{Witte1992,Court2003,Lu2010,Bell2010,Hopk2016}, which has a zero crossing in the magnetic field [Fig. \ref{fig:zeeman_spliting}(b)]. Although this approach has its merits, the literature \cite{Ali2016} implies that transverse field spin-flip slowers must be specially designed with transverse optical access in the zero-field region to allow for the necessary repumping and spin polarization \footnote{A special case of an efficient transverse field spin-flip slower was demonstrated with Sr \cite{Hill2014}, which has a $J=0$ lower cooling state and therefore does not suffer the depumping issues discussed in Ref. \cite{Ali2016}.}. Additionally, the indium spin flip slower still suffers from state mixing at the end of its length, which is a critical region for achieving well-slowed atomic beams. 

Finally, the third option is the decreasing field configuration \cite{Phil1982,Mole1997,Lison1999} (also referred as the ``$\sigma^+$ configuration''). In this approach, an unmixed transition can be driven over the entire length of the Zeeman slower [Fig. \ref{fig:zeeman_spliting}(c)]. Opting for this configuration, we choose $\Delta_0/2\pi = -220\ \unit{MHz}$, resulting in $B = 0$ at the end of Zeeman slower with exit velocity $v_f = 70\ \unit{m/s}$. Although a smaller $v_f$ may be beneficial, this would require $\Delta_0$ to be closer to resonance, and at small detunings atoms exiting the slower can scatter so many photons that they reverse direction and become untrappable \cite{Barr1991}.

\begin{figure}[htbp]
	\centering
	\includegraphics[width=\linewidth]{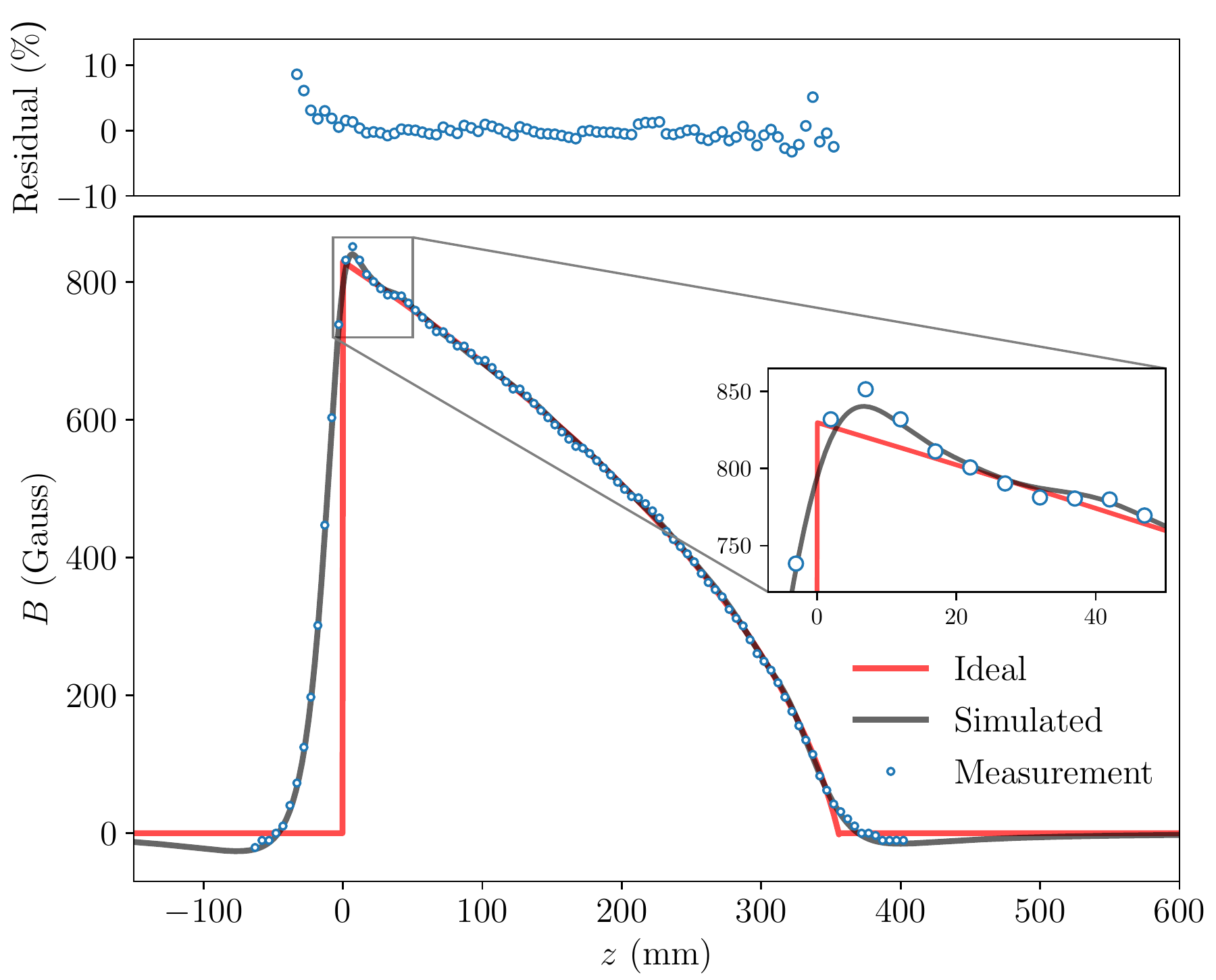}
	\caption{Top: residuals between the simulated field and measured field profiles for the region most relevant to slowing (i.e., when the simulated values are larger than 35 G). Residuals remain low except near high field gradients. Bottom: comparison between the ideal field (red curve), the numerically optimized field (black curve), and the measured field profiles (blue dots). Inset: field ripples at the beginning of the slower.}
	\label{fig:calculated_measured_field}
\end{figure}

To realize the desired field profile, we model our permanent magnets as magnetic dipoles \cite{Ovch2007,Hill2014}. Using 18 pairs of magnet stacks, we develop an iterative field optimization algorithm that varies the size of the magnets in each stack pair as well as their separation. The algorithm attempts to get the field of the magnet array as close as possible to the ideal field expression of Eq. (\ref{eq:ideal_B_field}) (Fig. \ref{fig:calculated_measured_field}). The optimized magnet design achieves close agreement with the desired field except for unavoidable discrepancies when the ideal model has sharp changes \cite{Chei2011,Hill2014, Ali2016}. To confirm that the numerically determined field is acceptable, we perform an atomic trajectory simulation \cite{Hopk2016} of atoms moving in our optimized permanent magnet field. The simulation suggests that efficient slowing is still possible with our design. Separate measurements with a magnetometer confirm that our calculated permanent magnet field is realized (Fig. \ref{fig:calculated_measured_field}).

\section{Zeeman slower performance}
The performance of our Zeeman slower is characterized by measuring the longitudinal velocity distribution of the atomic beam entering the science chamber (Fig. \ref{fig:apparatus}). These measurements are performed by collecting atom fluorescence as a function of the probe laser wavelength $\lambda_p$ (as determined by a precise HighFinesse wavemeter). Wavelength values are converted to velocities as $c \left(\frac{\lambda_0 - \lambda_p}{\lambda_p} \right) \sec(45^{\circ})$, where $\lambda_0$ is the zero-velocity cooling transition wavelength and $c$ is the speed of light. $\lambda_0$ is measured as the center of the cooling transition lineshape when the atoms are probed at a $\theta = 90^\circ$ angle with respect to the atomic beam (Fig. \ref{fig:apparatus}) \footnote{In the transverse direction, we observe a linewidth of 29 MHz, which is slightly broader than the 20.7 MHz natural linewidth.}. 

\begin{figure}[htbp]
	\centering
	\includegraphics[width=\linewidth]{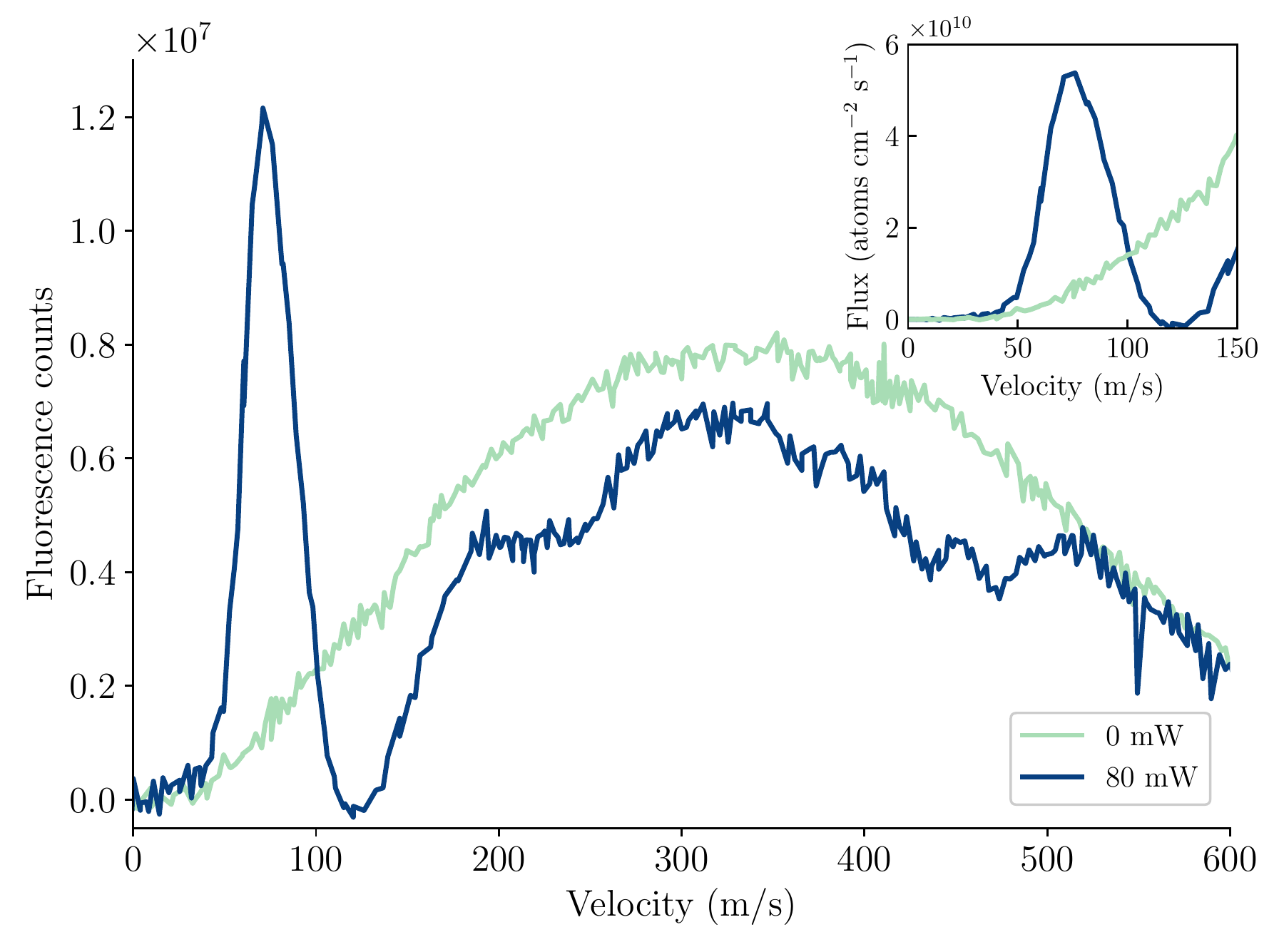}
	\caption{Longitudinal velocity distributions with and without the slower laser. The slowing laser detuning is kept constant at $\Delta_0/2\pi = -325\ \unit{MHz}$ during the measurement. Inset: the slowed atomic beam in flux units (Appendix \ref{sec:flux_calculation}).}
	\label{fig:full_range_data}
\end{figure}

Figure \ref{fig:full_range_data} shows the measured longitudinal velocity distribution under the influence of the Zeeman slower. For both curves, we use 12 mW per beam for both 410~nm state preparation lasers and 20 mW for the 326~nm spin polarization laser. For the repumpers, we use 30 mW for each of the 410~nm lasers and 23 mW for each of the 451~nm lasers. The 326~nm probe power is 490 $\unit{\mu W}$. The Zeeman slower takes effect around 500 m/s (similar to the $v_0 = 450\ \unit{m/s}$ design value). We observe a depletion of the distribution for velocities greater than 120 m/s, and we see a pronounced fluorescence peak near our design goal of $v_f = 70\ \unit{m/s}$ (at a slowing laser detuning of $\Delta_0/2\pi = -325\ \unit{MHz}$). Increasing the slowing laser beam waist would likely further slow the remaining fast atoms \cite{Wodey2021}. We note that the negative counts shown in the fluorescence measurements are an artifact of background subtraction (Appendix \ref{sec:background_subtraction}).

\begin{figure*}
\centering
\begin{minipage}[t]{.485\textwidth}
	\centering
	\includegraphics[width=\linewidth]{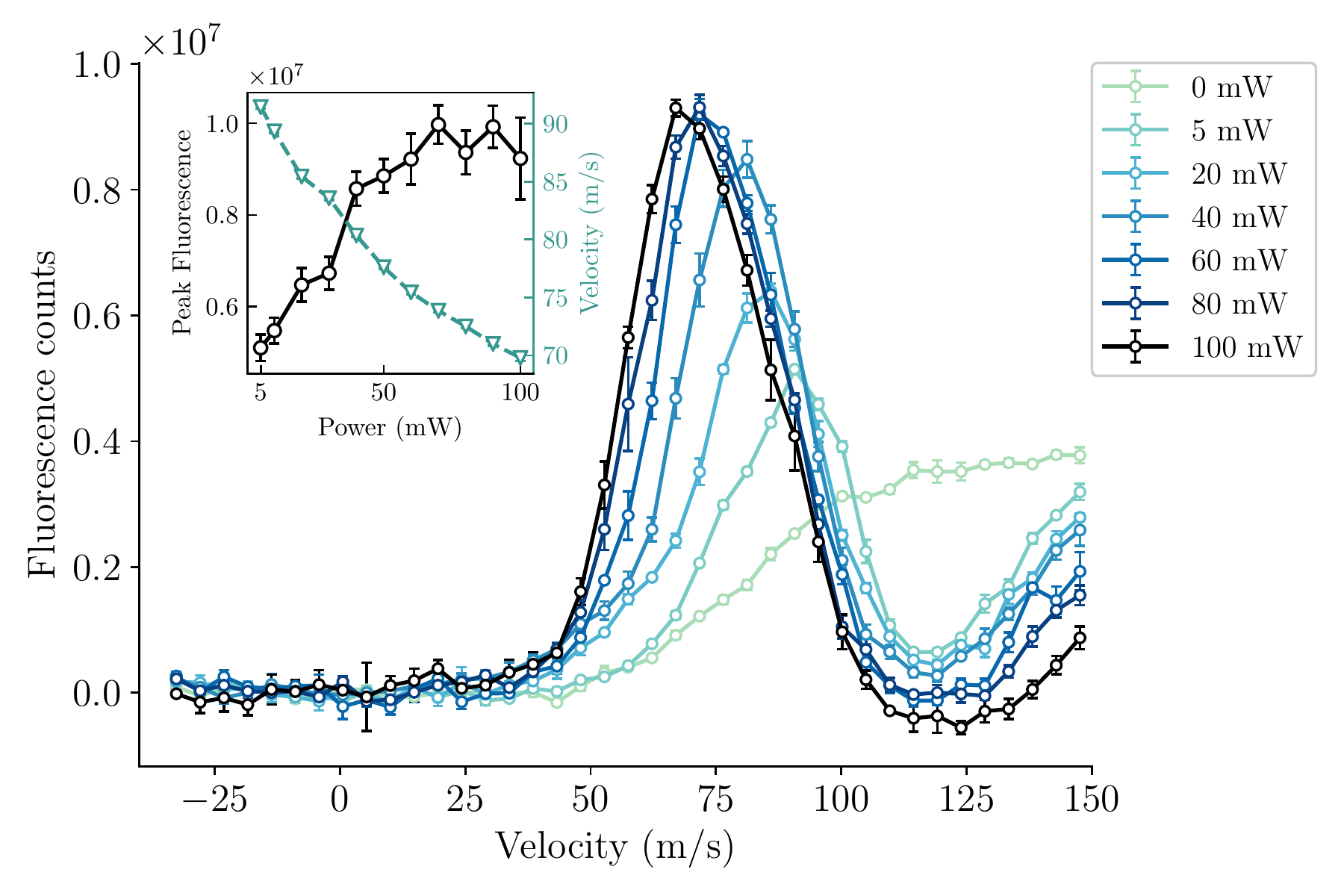}
	\caption{Slowing laser power dependence. The slowing laser detuning, state preparation laser power and repumper power are the same as in Fig. \ref{fig:full_range_data}. The science chamber probe beam power is 420 $\unit{\mu W}$. Inset: slowed atomic beam amplitudes and central velocities obtained by fitting the fluorescence with an asymmetric pseudo-Voigt profile \cite{Stan2008}. The trend indicates saturation above 70 mW.}
	\label{fig:slowing_beam_saturation}

\end{minipage}
\hfill
\begin{minipage}[t]{.485\textwidth}
	\centering
	\includegraphics[width=\linewidth]{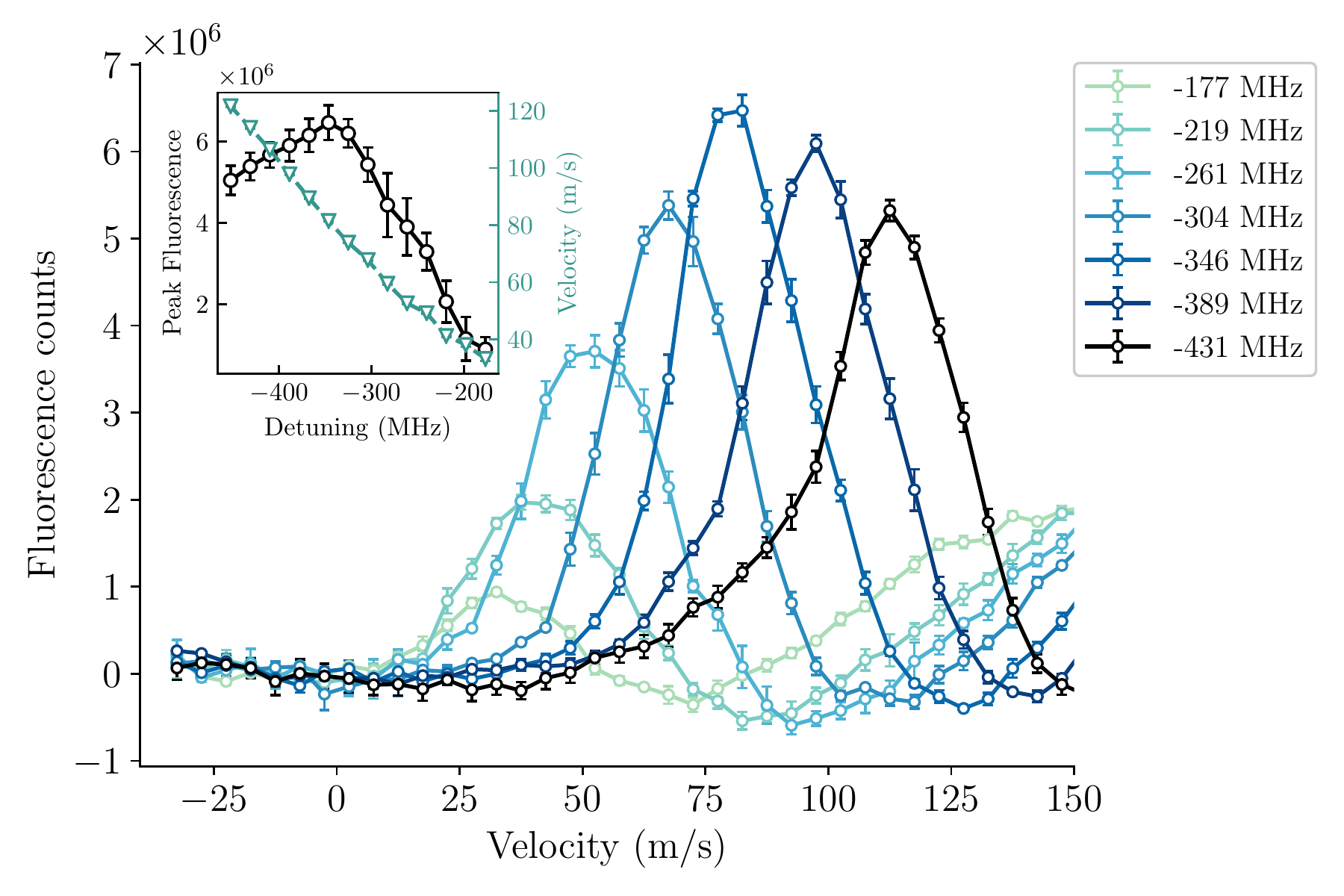}
	\caption{Slowing laser detuning dependence. Here the slowing laser power is fixed at 80 mW. The probe power is 480 $\unit{\mu W}$. Inset: fitting the peaks with an asymmetric pseudo-Voight profile, we find that the fluorescence maximizes at $\Delta_0/2\pi = -346\ \unit{MHz}$, where the slowed atomic beam peaks at $81.5\ \unit{m/s}$.}
	\label{fig:slowing_beam_detuning}

\end{minipage}
\end{figure*}

We have also characterized the slowed atomic beam as a function of slowing laser power (Fig. \ref{fig:slowing_beam_saturation}) and detuning (Fig. \ref{fig:slowing_beam_detuning}). We find that the slowing effect saturates above 70 mW (where $s=1.2$) and peaks at $\Delta_0/2\pi = -346\ \unit{MHz}$. For these measurements, we use the same state preparation and repumper laser powers as in the Fig. \ref{fig:full_range_data} data. One exception is the spin polarization laser, which is not used for the detuning characterization (Fig. \ref{fig:slowing_beam_detuning}) since it originates from the same source as the slowing laser and would not function properly if its detuning is varied. The detuning is set to be $\Delta_0/2\pi = -325\ \unit{MHz}$ for the power measurement, and the power is fixed at 80 mW for the detuning measurement. A $-325\ \unit{MHz}$ detuning is chosen because this is the where our design goal of $v_f = 70\ \unit{m/s}$ occurs. Although the fluorescence is better at $-346\ \unit{MHz}$, we only lose $\sim 4\%$ of the fluorescence amplitude by operating at $-325\ \unit{MHz}$.

To further optimize our Zeeman slower, we characterize our spin polarization and repumper lasers \footnote{A characterization of the 410~nm state preparation lasers is not described here because they do not affect our Zeeman slower efficiency (only the overall atom number)}. Atoms in the $\ket{5P_{3/2},F=6,m_F=6}$ state are most efficiently slowed; however, only applying two 410~nm lasers in the state preparation region would cause population to be distributed among all 13 $m_F$ states in $\ket{5P_{3/2},F=6}$. For this reason, the spin polarization laser improves slower performance. To define a quantization axis for spin polarization, we erect a pair of magnetic coils in the Helmholtz configuration, which provide a magnetic field of 3 G. The 326~nm spin polarization laser is slightly red detuned and circularly polarized, and it is aligned through the centers of the Helmholtz coils.

\begin{figure}[htbp]
	\centering
	\includegraphics[width=\linewidth]{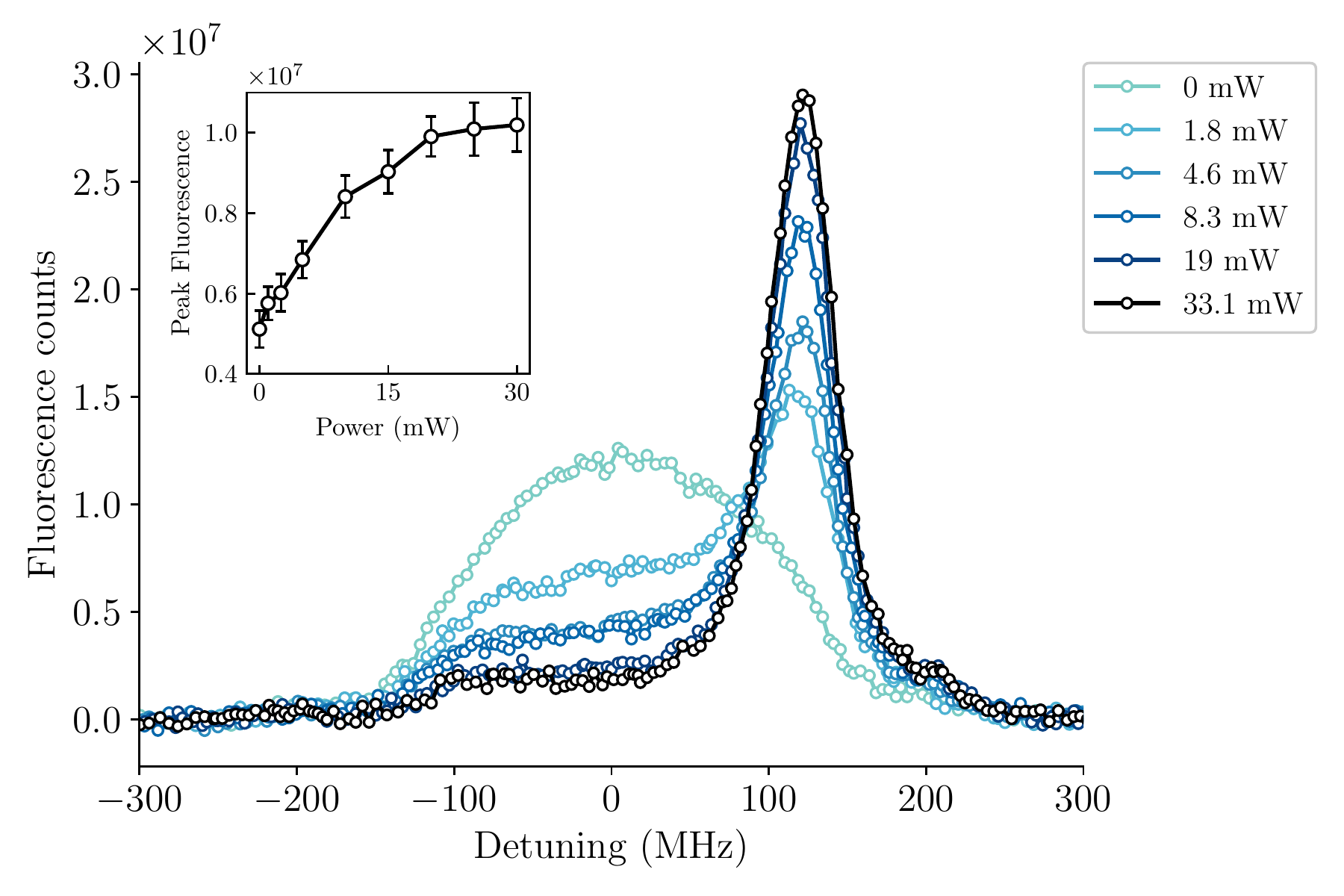}
	\caption{Fluorescence signals of spin polarized atoms with various spin polarization laser powers. These data were acquired in the science chamber in the presence of a 160 G bias field and without the use of the Zeeman slower. At maximum power, atoms are well polarized into $\ket{5P_{3/2},F=6,m_F=6}$. Inset: peak fluorescence of the slowed atomic beam as a function of spin polarization laser power. Here the system is run in its normal configuration, without the science chamber bias field and with the slower laser present. The effect saturates above 20 mW.}
	\label{fig:pol_power_saturation}
\end{figure}

Spin polarization is characterized by measuring the population in the Zeeman sublevels of the lower cooling state. For this measurement, the Zeeman slower laser is shut off, and we apply a 160 G bias field in the science chamber to split the sublevels. Meanwhile, the probe laser is aligned at $\theta = 90^\circ$ with respect to the atomic beam (Fig. \ref{fig:apparatus}), and its polarization is oriented along the bias field. This transverse probe alignment allows for better resolution of the sublevels since the transverse linewidths are much smaller than the longitudinal ones. We then scan the probe laser frequency to measure the population of the $\ket{5P_{3/2},F=6}$ Zeeman sublevels. Although the bias field is not large enough to fully resolve each $m_F$ state (Fig. \ref{fig:pol_power_saturation}), the Zeeman splitting is large enough for us to observe substantial polarization into the $\ket{5P_{3/2},F=6,m_F=6}$ state when the spin polarization laser power is above 20 mW (Fig. \ref{fig:pol_power_saturation}). 

To confirm the effect of spin polarization on the slowed atomic beam, we restore our system to its normal configuration (i.e., no science chamber bias field, the optimized slower laser present, and the probe at $\theta = 45^\circ$). We find that above 20 mW, the spin polarization laser doubles the output of our Zeeman slower (Fig. \ref{fig:pol_power_saturation} inset). 

The slower performance also depends on repump detuning. This is because the value of $\mu_{\textit{eff}}$ for the repumper transitions is different than that of the cooling transition, so the repumpers cannot be resonant with all velocities throughout the slower \cite{Chei2011}. We determine the repumper detunings empirically by using them to maximize the slowed atomic beam fluorescence. The optimal values are shown in Table \ref{tab:repump_detunings}. The uncertainties in these values reflect the range over which there was no change in the fluorescence signal. It has been suggested that insensitivity to Zeeman slower repumper frequency is the result of multiple leaking mechanisms along the slower \cite{Chei2011}.

\begin{table}
\caption{Optimal detunings for each repumper transition}
\begin{ruledtabular}
\begin{tabular}{c c}
    Transition & Detuning (MHz)\\ 
    \hline
    $\ket{5P_{1/2},F=4} \rightarrow \ket{6S_{1/2},F=5}$ & $-300\pm40$\\  
    $\ket{5P_{1/2},F=5} \rightarrow \ket{6S_{1/2},F=5}$ & $-260\pm40$\\
    $\ket{5P_{3/2},F=4} \rightarrow \ket{6S_{1/2},F=5}$ & $-450\pm30$\\  
    $\ket{5P_{3/2},F=5} \rightarrow \ket{6S_{1/2},F=5}$ & $-530\pm30$\\
\end{tabular}
\end{ruledtabular}
\label{tab:repump_detunings}
\end{table}

\begin{figure}[htbp]
	\centering
	\includegraphics[width=\linewidth]{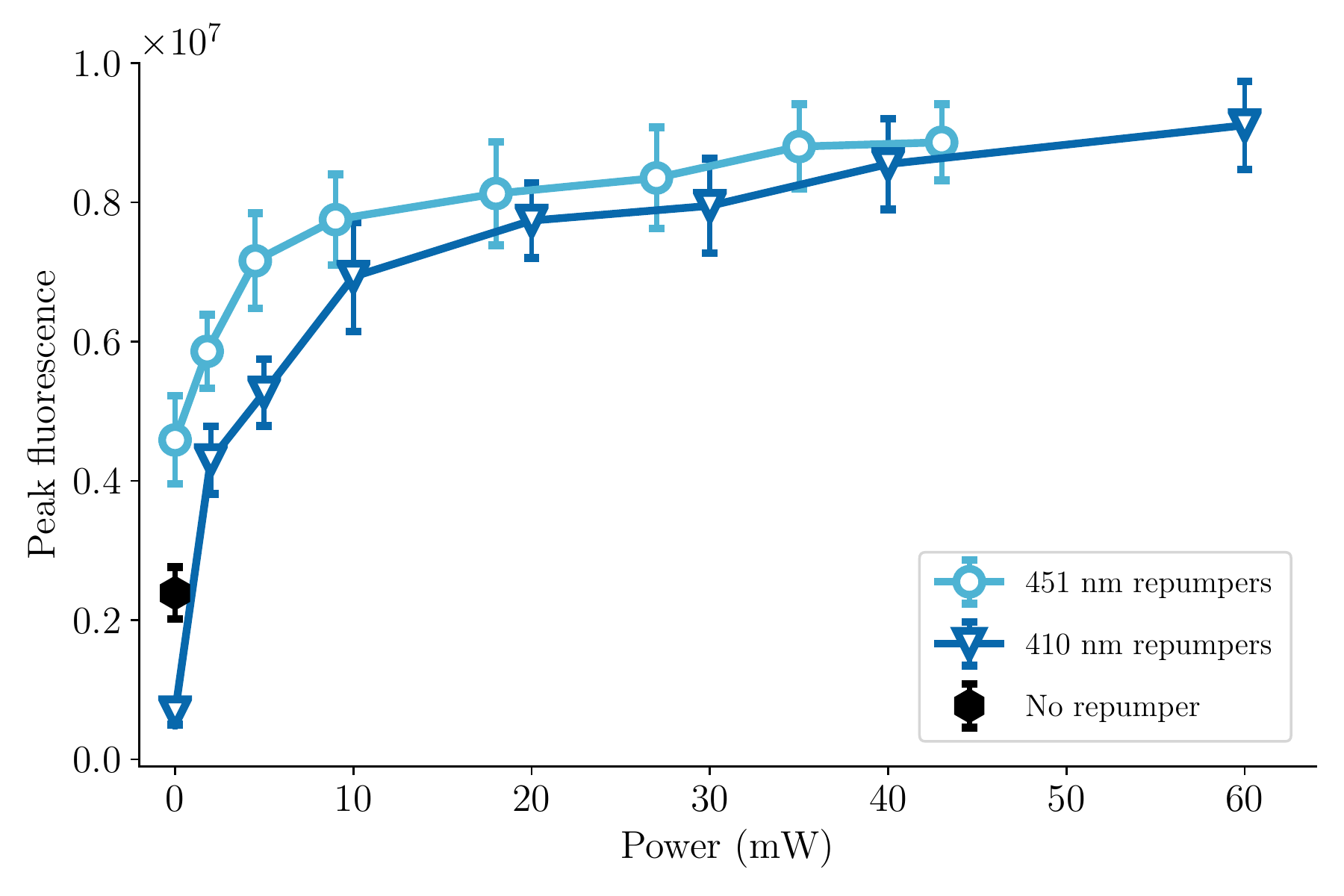}
	\caption{Slowed atomic beam fluorescence amplitude as a function of repumper power. The $x$\_axis represents the combined power for two repumper beams of a given wavelength. Here the slowing laser power is fixed at 80 mW and its detuning is set to be $-325\ \unit{MHz}$. The probe beam power is $420\ \unit{\mu W}$. Compared to no repumping, the slowed atomic beam brightness increases by a factor of 5 with the repumper lasers.}
	\label{fig:repumper_power_saturation}
\end{figure}

The saturation behavior of the repumpers is depicted in Fig. \ref{fig:repumper_power_saturation}. Here we vary the 410~nm (451~nm) laser power while fixing the 451~nm (410~nm) lasers at maximum power. The $x$\_axis represents the combined power of the two 410~nm or 451~nm lasers. For the 451~nm repumpers, the peak fluorescence of the slowed atomic beam saturates when the power reaches 30 mW. For the 410~nm lasers, modest gains still seem possible beyond our maximum power of 60 mW. Higher power 410~nm lasers might provide a marginal benefit and frequency broadened 410~nm lasers might also substantially improve repumper efficiency \cite{Chei2011,Petz2018}. With ample repumper power, the slowed atomic beam is five times brighter than in the absence of repumpers.

\section{Summary and outlook}
We demonstrate a slowed atomic beam of a triel atom. Our setup is a transverse permanent magnet Zeeman slower in decreasing field configuration. We characterize slowing, state preparation, and repumping, and we ultimately achieve a $70\ \unit{m/s}$ final velocity. Further improvements are possible with larger slowing laser waists (to target more of the atomic beam), modulating repumper frequencies \cite{Chei2011,Petz2018}, and implementing a longitudinal field slower to reduce the required slowing laser power \cite{Hill2014}. Furthermore, a blue-detuned slower might allow for smaller final velocities \cite{Lison1999}. Our work extends to all atoms with similar energy structure, such as those in main group III of the Periodic Table. This opens up possibilities for laser cooling and trapping of triel atoms, which are unexplored in the ultracold regime.

\begin{acknowledgments}
This work was supported by the National Research Foundation, Prime Minister’s Office, Singapore and the Ministry of Education, Singapore under the Research Centres of Excellence program.
\end{acknowledgments}

\appendix
\section*{Appendix: DATA PROCESSING}

\subsection{Negative fluorescence counts}
\label{sec:background_subtraction}
Figures \ref{fig:full_range_data}, \ref{fig:slowing_beam_saturation}, and \ref{fig:slowing_beam_detuning} depict negative values for fluorescence counts, which is due to background subtraction. The background levels for each measured curve in the three figures is taken to be the counts recorded at zero velocity where the probe and slower overlap. The main contributor to this background level is scattering of the slower laser from higher-velocity particles in the atomic beam.

When the probe and slower laser have equal frequencies, we observe a depletion of the background counts below our subtracted level, resulting in the negative fluorescence regimes in Figs. \ref{fig:full_range_data}, \ref{fig:slowing_beam_saturation}, and \ref{fig:slowing_beam_detuning}. We attribute this depletion to a small amount of additional slowing of the atomic beam. Before atoms enter the permanent magnet array and after they emerge from it, those that are Doppler shifted into resonance with the slower laser will experience a bit of additional slowing due to scattering slower laser photons.

\subsection{Flux calculation}
\label{sec:flux_calculation}
The inset of Fig. \ref{fig:full_range_data} shows the slowed atomic beam in flux units. The flux $\mathcal{F}$ is given by $\mathcal{F} = n v$, where $n$ is the atomic density. The density is extracted from fitting the distribution of probe fluorescence counts when the probe is resonant with the slowed atomic beam. The fit provides the physical dimensions of the volume of overlap between the probe and the atomic beam. 

The atomic density is given by

\begin{equation}
    n = \frac{N_{counts}}{QE \, t_{ex} P_{\Omega} {\Gamma_s} V},
\end{equation}
where $N_{counts}$ is the number of camera counts measured in the overlap volume, $QE$ is the quantum efficiency of the camera, $t_{ex}$ is the camera exposure time, $\Gamma_s = \frac{\Gamma}{2}\frac{s}{1+s}$ is the single-atom scattering rate, and $V$ is the overlap volume obtained from the fluorescence distribution fit. 

The angular emission probability $P_{\Omega}$ is determined as follows. The probe laser polarization is perpendicular to the camera imaging axis to maximize dipole emission into the camera, so the angular probability distribution of dipole emission \cite{steck2021} is

\begin{equation}
    f_{\Omega}(\theta,\phi) = \frac{3}{8\pi} \left( 1 - \sin^2\theta \cos^2\phi \right).
\end{equation}
Here we choose a coordinate system where the imaging axis points in the $z$ direction and the probe polarization is along $x$, resulting in the polar and azimuthal angles $\theta$ and $\phi$ (note that this $z$ axis is different than the one in the main text). The angle subtended by the imaging system is $\alpha = \tan^{-1}(D/4f)$, where $D$ is the imaging lens diameter and $f$ is the focal length. Therefore,

\begin{align}
    P_{\Omega} &= \int_0^{2\pi} d\phi \int_0^\alpha d\theta \sin\theta \, f_{\Omega}(\theta,\phi) \\
    &= \frac12 - \frac38 \cos\alpha - \frac18 \cos^3\alpha \\
    &\simeq \frac38 \left( \frac{D}{4f} \right)^2 \mbox{(when $D<<4f$)}.
\end{align}

\bibliography{references.bib}

\begin{thebibliography}{47}%
\makeatletter
\providecommand \@ifxundefined [1]{%
 \@ifx{#1\undefined}
}%
\providecommand \@ifnum [1]{%
 \ifnum #1\expandafter \@firstoftwo
 \else \expandafter \@secondoftwo
 \fi
}%
\providecommand \@ifx [1]{%
 \ifx #1\expandafter \@firstoftwo
 \else \expandafter \@secondoftwo
 \fi
}%
\providecommand \natexlab [1]{#1}%
\providecommand \enquote  [1]{``#1''}%
\providecommand \bibnamefont  [1]{#1}%
\providecommand \bibfnamefont [1]{#1}%
\providecommand \citenamefont [1]{#1}%
\providecommand \href@noop [0]{\@secondoftwo}%
\providecommand \href [0]{\begingroup \@sanitize@url \@href}%
\providecommand \@href[1]{\@@startlink{#1}\@@href}%
\providecommand \@@href[1]{\endgroup#1\@@endlink}%
\providecommand \@sanitize@url [0]{\catcode `\\12\catcode `\$12\catcode
  `\&12\catcode `\#12\catcode `\^12\catcode `\_12\catcode `\%12\relax}%
\providecommand \@@startlink[1]{}%
\providecommand \@@endlink[0]{}%
\providecommand \url  [0]{\begingroup\@sanitize@url \@url }%
\providecommand \@url [1]{\endgroup\@href {#1}{\urlprefix }}%
\providecommand \urlprefix  [0]{URL }%
\providecommand \Eprint [0]{\href }%
\providecommand \doibase [0]{https://doi.org/}%
\providecommand \selectlanguage [0]{\@gobble}%
\providecommand \bibinfo  [0]{\@secondoftwo}%
\providecommand \bibfield  [0]{\@secondoftwo}%
\providecommand \translation [1]{[#1]}%
\providecommand \BibitemOpen [0]{}%
\providecommand \bibitemStop [0]{}%
\providecommand \bibitemNoStop [0]{.\EOS\space}%
\providecommand \EOS [0]{\spacefactor3000\relax}%
\providecommand \BibitemShut  [1]{\csname bibitem#1\endcsname}%
\let\auto@bib@innerbib\@empty
\bibitem [{\citenamefont {Schreck}\ and\ \citenamefont {van
  Druten}(2021)}]{Schr2021}%
  \BibitemOpen
  \bibfield  {author} {\bibinfo {author} {\bibfnamefont {F.}~\bibnamefont
  {Schreck}}\ and\ \bibinfo {author} {\bibfnamefont {K.}~\bibnamefont {van
  Druten}},\ }\bibfield  {title} {\bibinfo {title} {Laser cooling for quantum
  gases},\ }\href@noop {} {\bibfield  {journal} {\bibinfo  {journal} {Nature
  Physics}\ }\textbf {\bibinfo {volume} {17}},\ \bibinfo {pages} {1296}
  (\bibinfo {year} {2021})}\BibitemShut {NoStop}%
\bibitem [{\citenamefont {McGowan}\ \emph {et~al.}(1995)\citenamefont
  {McGowan}, \citenamefont {Giltner},\ and\ \citenamefont {Lee}}]{McGowan95}%
  \BibitemOpen
  \bibfield  {author} {\bibinfo {author} {\bibfnamefont {R.~W.}\ \bibnamefont
  {McGowan}}, \bibinfo {author} {\bibfnamefont {D.~M.}\ \bibnamefont
  {Giltner}},\ and\ \bibinfo {author} {\bibfnamefont {S.~A.}\ \bibnamefont
  {Lee}},\ }\bibfield  {title} {\bibinfo {title} {Light force cooling,
  focusing, and nanometer-scale deposition of aluminum atoms},\ }\href@noop {}
  {\bibfield  {journal} {\bibinfo  {journal} {Opt. Lett.}\ }\textbf {\bibinfo
  {volume} {20}},\ \bibinfo {pages} {2535} (\bibinfo {year}
  {1995})}\BibitemShut {NoStop}%
\bibitem [{\citenamefont {Rehse}\ \emph {et~al.}(2004)\citenamefont {Rehse},
  \citenamefont {Bockel},\ and\ \citenamefont {Lee}}]{Rehse2004}%
  \BibitemOpen
  \bibfield  {author} {\bibinfo {author} {\bibfnamefont {S.~J.}\ \bibnamefont
  {Rehse}}, \bibinfo {author} {\bibfnamefont {K.~M.}\ \bibnamefont {Bockel}},\
  and\ \bibinfo {author} {\bibfnamefont {S.~A.}\ \bibnamefont {Lee}},\
  }\bibfield  {title} {\bibinfo {title} {Laser collimation of an atomic gallium
  beam},\ }\href@noop {} {\bibfield  {journal} {\bibinfo  {journal} {Phys. Rev.
  A}\ }\textbf {\bibinfo {volume} {69}},\ \bibinfo {pages} {063404} (\bibinfo
  {year} {2004})}\BibitemShut {NoStop}%
\bibitem [{\citenamefont {Kl\"oter}\ \emph {et~al.}(2008)\citenamefont
  {Kl\"oter}, \citenamefont {Weber}, \citenamefont {Haubrich}, \citenamefont
  {Meschede},\ and\ \citenamefont {Metcalf}}]{Kloter2008}%
  \BibitemOpen
  \bibfield  {author} {\bibinfo {author} {\bibfnamefont {B.}~\bibnamefont
  {Kl\"oter}}, \bibinfo {author} {\bibfnamefont {C.}~\bibnamefont {Weber}},
  \bibinfo {author} {\bibfnamefont {D.}~\bibnamefont {Haubrich}}, \bibinfo
  {author} {\bibfnamefont {D.}~\bibnamefont {Meschede}},\ and\ \bibinfo
  {author} {\bibfnamefont {H.}~\bibnamefont {Metcalf}},\ }\bibfield  {title}
  {\bibinfo {title} {Laser cooling of an indium atomic beam enabled by magnetic
  fields},\ }\href@noop {} {\bibfield  {journal} {\bibinfo  {journal} {Phys.
  Rev. A}\ }\textbf {\bibinfo {volume} {77}},\ \bibinfo {pages} {033402}
  (\bibinfo {year} {2008})}\BibitemShut {NoStop}%
\bibitem [{\citenamefont {Kim}\ \emph {et~al.}(2009)\citenamefont {Kim},
  \citenamefont {Haubrich},\ and\ \citenamefont {Meschede}}]{Kim2009}%
  \BibitemOpen
  \bibfield  {author} {\bibinfo {author} {\bibfnamefont {J.-I.}\ \bibnamefont
  {Kim}}, \bibinfo {author} {\bibfnamefont {D.}~\bibnamefont {Haubrich}},\ and\
  \bibinfo {author} {\bibfnamefont {D.}~\bibnamefont {Meschede}},\ }\bibfield
  {title} {\bibinfo {title} {Efficient sub-doppler laser cooling of an indium
  atomic beam},\ }\href@noop {} {\bibfield  {journal} {\bibinfo  {journal}
  {Opt. Express}\ }\textbf {\bibinfo {volume} {17}},\ \bibinfo {pages} {21216}
  (\bibinfo {year} {2009})}\BibitemShut {NoStop}%
\bibitem [{\citenamefont {Fan}\ \emph {et~al.}(2011)\citenamefont {Fan},
  \citenamefont {Chen}, \citenamefont {Liu}, \citenamefont {Lien},
  \citenamefont {Shy},\ and\ \citenamefont {Liu}}]{Fan2011}%
  \BibitemOpen
  \bibfield  {author} {\bibinfo {author} {\bibfnamefont {I.}~\bibnamefont
  {Fan}}, \bibinfo {author} {\bibfnamefont {T.-L.}\ \bibnamefont {Chen}},
  \bibinfo {author} {\bibfnamefont {Y.-S.}\ \bibnamefont {Liu}}, \bibinfo
  {author} {\bibfnamefont {Y.-H.}\ \bibnamefont {Lien}}, \bibinfo {author}
  {\bibfnamefont {J.-T.}\ \bibnamefont {Shy}},\ and\ \bibinfo {author}
  {\bibfnamefont {Y.-W.}\ \bibnamefont {Liu}},\ }\bibfield  {title} {\bibinfo
  {title} {Prospects of laser cooling in atomic thallium},\ }\href@noop {}
  {\bibfield  {journal} {\bibinfo  {journal} {Phys. Rev. A}\ }\textbf {\bibinfo
  {volume} {84}},\ \bibinfo {pages} {042504} (\bibinfo {year}
  {2011})}\BibitemShut {NoStop}%
\bibitem [{\citenamefont {Phillips}\ and\ \citenamefont
  {Metcalf}(1982)}]{Phil1982}%
  \BibitemOpen
  \bibfield  {author} {\bibinfo {author} {\bibfnamefont {W.~D.}\ \bibnamefont
  {Phillips}}\ and\ \bibinfo {author} {\bibfnamefont {H.}~\bibnamefont
  {Metcalf}},\ }\bibfield  {title} {\bibinfo {title} {Laser deceleration of an
  atomic beam},\ }\href@noop {} {\bibfield  {journal} {\bibinfo  {journal}
  {Phys. Rev. Lett.}\ }\textbf {\bibinfo {volume} {48}},\ \bibinfo {pages}
  {596} (\bibinfo {year} {1982})}\BibitemShut {NoStop}%
\bibitem [{\citenamefont {Dieckmann}\ \emph {et~al.}(1998)\citenamefont
  {Dieckmann}, \citenamefont {Spreeuw}, \citenamefont {Weidem{\"u}ller},\ and\
  \citenamefont {Walraven}}]{dieckmann1998}%
  \BibitemOpen
  \bibfield  {author} {\bibinfo {author} {\bibfnamefont {K.}~\bibnamefont
  {Dieckmann}}, \bibinfo {author} {\bibfnamefont {R.}~\bibnamefont {Spreeuw}},
  \bibinfo {author} {\bibfnamefont {M.}~\bibnamefont {Weidem{\"u}ller}},\ and\
  \bibinfo {author} {\bibfnamefont {J.}~\bibnamefont {Walraven}},\ }\bibfield
  {title} {\bibinfo {title} {Two-dimensional magneto-optical trap as a source
  of slow atoms},\ }\href@noop {} {\bibfield  {journal} {\bibinfo  {journal}
  {Phys. Rev. A}\ }\textbf {\bibinfo {volume} {58}},\ \bibinfo {pages} {3891}
  (\bibinfo {year} {1998})}\BibitemShut {NoStop}%
\bibitem [{\citenamefont {Lamporesi}\ \emph {et~al.}(2013)\citenamefont
  {Lamporesi}, \citenamefont {Donadello}, \citenamefont {Serafini},\ and\
  \citenamefont {Ferrari}}]{lamporesi2013}%
  \BibitemOpen
  \bibfield  {author} {\bibinfo {author} {\bibfnamefont {G.}~\bibnamefont
  {Lamporesi}}, \bibinfo {author} {\bibfnamefont {S.}~\bibnamefont
  {Donadello}}, \bibinfo {author} {\bibfnamefont {S.}~\bibnamefont
  {Serafini}},\ and\ \bibinfo {author} {\bibfnamefont {G.}~\bibnamefont
  {Ferrari}},\ }\bibfield  {title} {\bibinfo {title} {Compact high-flux source
  of cold sodium atoms},\ }\href@noop {} {\bibfield  {journal} {\bibinfo
  {journal} {Rev. Sci. Instrum.}\ }\textbf {\bibinfo {volume} {84}},\ \bibinfo
  {pages} {063102} (\bibinfo {year} {2013})}\BibitemShut {NoStop}%
\bibitem [{\citenamefont {Ovchinnikov}(2007)}]{Ovch2007}%
  \BibitemOpen
  \bibfield  {author} {\bibinfo {author} {\bibfnamefont {Y.~B.}\ \bibnamefont
  {Ovchinnikov}},\ }\bibfield  {title} {\bibinfo {title} {A zeeman slower based
  on magnetic dipoles},\ }\href@noop {} {\bibfield  {journal} {\bibinfo
  {journal} {Optics Communications}\ }\textbf {\bibinfo {volume} {276}},\
  \bibinfo {pages} {261} (\bibinfo {year} {2007})}\BibitemShut {NoStop}%
\bibitem [{\citenamefont {Eck}\ \emph {et~al.}(1957)\citenamefont {Eck},
  \citenamefont {Lurio},\ and\ \citenamefont {Kusch}}]{Eck1957}%
  \BibitemOpen
  \bibfield  {author} {\bibinfo {author} {\bibfnamefont {T.~G.}\ \bibnamefont
  {Eck}}, \bibinfo {author} {\bibfnamefont {A.}~\bibnamefont {Lurio}},\ and\
  \bibinfo {author} {\bibfnamefont {P.}~\bibnamefont {Kusch}},\ }\bibfield
  {title} {\bibinfo {title} {Hfs of the $5^{2}p_{\frac{1}{2}}$ state of
  {$\mathrm{In}^{115}$} and {$\mathrm{In}^{113}$}: Hfs anomalies in the stable
  isotopes of indium},\ }\href@noop {} {\bibfield  {journal} {\bibinfo
  {journal} {Phys. Rev.}\ }\textbf {\bibinfo {volume} {106}},\ \bibinfo {pages}
  {954} (\bibinfo {year} {1957})}\BibitemShut {NoStop}%
\bibitem [{\citenamefont {Gunawardena}\ \emph {et~al.}(2009)\citenamefont
  {Gunawardena}, \citenamefont {Cao}, \citenamefont {Hess},\ and\ \citenamefont
  {Majumder}}]{Gunawardena2009}%
  \BibitemOpen
  \bibfield  {author} {\bibinfo {author} {\bibfnamefont {M.}~\bibnamefont
  {Gunawardena}}, \bibinfo {author} {\bibfnamefont {H.}~\bibnamefont {Cao}},
  \bibinfo {author} {\bibfnamefont {P.~W.}\ \bibnamefont {Hess}},\ and\
  \bibinfo {author} {\bibfnamefont {P.~K.}\ \bibnamefont {Majumder}},\
  }\bibfield  {title} {\bibinfo {title} {Measurement of hyperfine structure
  within the $6p_{3/2}$ excited state of $^{115}\mathrm{In}$},\ }\href@noop {}
  {\bibfield  {journal} {\bibinfo  {journal} {Phys. Rev. A}\ }\textbf {\bibinfo
  {volume} {80}},\ \bibinfo {pages} {032519} (\bibinfo {year}
  {2009})}\BibitemShut {NoStop}%
\bibitem [{\citenamefont {Zimmermann}(1970)}]{Zimmermann1970}%
  \BibitemOpen
  \bibfield  {author} {\bibinfo {author} {\bibfnamefont {P.}~\bibnamefont
  {Zimmermann}},\ }\bibfield  {title} {\bibinfo {title}
  {Level-crossing-experimente zur untersuchung der hyperfeinstruktur des
  {$5d^{2}D_{5/2}$}-terms im indium i-spektrum},\ }\href@noop {} {\bibfield
  {journal} {\bibinfo  {journal} {Zeitschrift f{\"u}r Physik}\ }\textbf
  {\bibinfo {volume} {233}},\ \bibinfo {pages} {21} (\bibinfo {year}
  {1970})}\BibitemShut {NoStop}%
\bibitem [{\citenamefont {Safronova}\ \emph {et~al.}(2007)\citenamefont
  {Safronova}, \citenamefont {Safronova},\ and\ \citenamefont
  {Kozlov}}]{Safr2007}%
  \BibitemOpen
  \bibfield  {author} {\bibinfo {author} {\bibfnamefont {U.~I.}\ \bibnamefont
  {Safronova}}, \bibinfo {author} {\bibfnamefont {M.~S.}\ \bibnamefont
  {Safronova}},\ and\ \bibinfo {author} {\bibfnamefont {M.~G.}\ \bibnamefont
  {Kozlov}},\ }\bibfield  {title} {\bibinfo {title} {Relativistic all-order
  calculations of {In I} and {Si II} atomic properties},\ }\href@noop {}
  {\bibfield  {journal} {\bibinfo  {journal} {Phys. Rev. A}\ }\textbf {\bibinfo
  {volume} {76}},\ \bibinfo {pages} {022501} (\bibinfo {year}
  {2007})}\BibitemShut {NoStop}%
\bibitem [{\citenamefont {Sahoo}\ and\ \citenamefont {Das}(2011)}]{Sahoo2011}%
  \BibitemOpen
  \bibfield  {author} {\bibinfo {author} {\bibfnamefont {B.~K.}\ \bibnamefont
  {Sahoo}}\ and\ \bibinfo {author} {\bibfnamefont {B.~P.}\ \bibnamefont
  {Das}},\ }\bibfield  {title} {\bibinfo {title} {Transition properties of
  low-lying states in atomic indium},\ }\href@noop {} {\bibfield  {journal}
  {\bibinfo  {journal} {Phys. Rev. A}\ }\textbf {\bibinfo {volume} {84}},\
  \bibinfo {pages} {012501} (\bibinfo {year} {2011})}\BibitemShut {NoStop}%
\bibitem [{\citenamefont {Ali}\ \emph {et~al.}(2017)\citenamefont {Ali},
  \citenamefont {Badr}, \citenamefont {Br{\'{e}}zillon}, \citenamefont
  {Dubessy}, \citenamefont {Perrin},\ and\ \citenamefont {Perrin}}]{Ali2017}%
  \BibitemOpen
  \bibfield  {author} {\bibinfo {author} {\bibfnamefont {D.~B.}\ \bibnamefont
  {Ali}}, \bibinfo {author} {\bibfnamefont {T.}~\bibnamefont {Badr}}, \bibinfo
  {author} {\bibfnamefont {T.}~\bibnamefont {Br{\'{e}}zillon}}, \bibinfo
  {author} {\bibfnamefont {R.}~\bibnamefont {Dubessy}}, \bibinfo {author}
  {\bibfnamefont {H.}~\bibnamefont {Perrin}},\ and\ \bibinfo {author}
  {\bibfnamefont {A.}~\bibnamefont {Perrin}},\ }\bibfield  {title} {\bibinfo
  {title} {Detailed study of a transverse field zeeman slower},\ }\href@noop {}
  {\bibfield  {journal} {\bibinfo  {journal} {Journal of Physics B: Atomic,
  Molecular and Optical Physics}\ }\textbf {\bibinfo {volume} {50}},\ \bibinfo
  {pages} {055008} (\bibinfo {year} {2017})}\BibitemShut {NoStop}%
\bibitem [{\citenamefont {Lison}\ \emph {et~al.}(1999)\citenamefont {Lison},
  \citenamefont {Schuh}, \citenamefont {Haubrich},\ and\ \citenamefont
  {Meschede}}]{Lison1999}%
  \BibitemOpen
  \bibfield  {author} {\bibinfo {author} {\bibfnamefont {F.}~\bibnamefont
  {Lison}}, \bibinfo {author} {\bibfnamefont {P.}~\bibnamefont {Schuh}},
  \bibinfo {author} {\bibfnamefont {D.}~\bibnamefont {Haubrich}},\ and\
  \bibinfo {author} {\bibfnamefont {D.}~\bibnamefont {Meschede}},\ }\bibfield
  {title} {\bibinfo {title} {High-brilliance zeeman-slowed cesium atomic
  beam},\ }\href@noop {} {\bibfield  {journal} {\bibinfo  {journal} {Phys. Rev.
  A}\ }\textbf {\bibinfo {volume} {61}},\ \bibinfo {pages} {013405} (\bibinfo
  {year} {1999})}\BibitemShut {NoStop}%
\bibitem [{\citenamefont {Reinaudi}\ \emph {et~al.}(2012)\citenamefont
  {Reinaudi}, \citenamefont {Osborn}, \citenamefont {Bega},\ and\ \citenamefont
  {Zelevinsky}}]{Rein2012}%
  \BibitemOpen
  \bibfield  {author} {\bibinfo {author} {\bibfnamefont {G.}~\bibnamefont
  {Reinaudi}}, \bibinfo {author} {\bibfnamefont {C.~B.}\ \bibnamefont
  {Osborn}}, \bibinfo {author} {\bibfnamefont {K.}~\bibnamefont {Bega}},\ and\
  \bibinfo {author} {\bibfnamefont {T.}~\bibnamefont {Zelevinsky}},\ }\bibfield
   {title} {\bibinfo {title} {Dynamically configurable and optimizable zeeman
  slower using permanent magnets and servomotors},\ }\href@noop {} {\bibfield
  {journal} {\bibinfo  {journal} {J. Opt. Soc. Am. B}\ }\textbf {\bibinfo
  {volume} {29}},\ \bibinfo {pages} {729} (\bibinfo {year} {2012})}\BibitemShut
  {NoStop}%
\bibitem [{\citenamefont {Hill}\ \emph {et~al.}(2014)\citenamefont {Hill},
  \citenamefont {Ovchinnikov}, \citenamefont {Bridge}, \citenamefont {Curtis},\
  and\ \citenamefont {Gill}}]{Hill2014}%
  \BibitemOpen
  \bibfield  {author} {\bibinfo {author} {\bibfnamefont {I.~R.}\ \bibnamefont
  {Hill}}, \bibinfo {author} {\bibfnamefont {Y.~B.}\ \bibnamefont
  {Ovchinnikov}}, \bibinfo {author} {\bibfnamefont {E.~M.}\ \bibnamefont
  {Bridge}}, \bibinfo {author} {\bibfnamefont {E.~A.}\ \bibnamefont {Curtis}},\
  and\ \bibinfo {author} {\bibfnamefont {P.}~\bibnamefont {Gill}},\ }\bibfield
  {title} {\bibinfo {title} {Zeeman slowers for strontium based on permanent
  magnets},\ }\href@noop {} {\bibfield  {journal} {\bibinfo  {journal} {Journal
  of Physics B: Atomic, Molecular and Optical Physics}\ }\textbf {\bibinfo
  {volume} {47}},\ \bibinfo {pages} {075006} (\bibinfo {year}
  {2014})}\BibitemShut {NoStop}%
\bibitem [{\citenamefont {Cheiney}(2013)}]{Chei2013}%
  \BibitemOpen
  \bibfield  {author} {\bibinfo {author} {\bibfnamefont {P.}~\bibnamefont
  {Cheiney}},\ }\emph {\bibinfo {title} {{Matter wave scattering on complex
  potentials}}},\ \href@noop {} {\bibinfo {type} {Theses}},\ \bibinfo  {school}
  {{Universit{\'e} Paul Sabatier - Toulouse III}} (\bibinfo {year}
  {2013})\BibitemShut {NoStop}%
\bibitem [{\citenamefont {Melentiev}\ \emph {et~al.}(2004)\citenamefont
  {Melentiev}, \citenamefont {Borisov},\ and\ \citenamefont
  {Balykin}}]{Mele2004}%
  \BibitemOpen
  \bibfield  {author} {\bibinfo {author} {\bibfnamefont {P.~N.}\ \bibnamefont
  {Melentiev}}, \bibinfo {author} {\bibfnamefont {P.~A.}\ \bibnamefont
  {Borisov}},\ and\ \bibinfo {author} {\bibfnamefont {V.~I.}\ \bibnamefont
  {Balykin}},\ }\bibfield  {title} {\bibinfo {title} {Zeeman laser cooling of
  {$^{85}\mathrm{Rb}$} atoms in transverse magnetic field},\ }\href@noop {}
  {\bibfield  {journal} {\bibinfo  {journal} {Journal of Experimental and
  Theoretical Physics}\ }\textbf {\bibinfo {volume} {98}},\ \bibinfo {pages}
  {667} (\bibinfo {year} {2004})}\BibitemShut {NoStop}%
\bibitem [{\citenamefont {Hopkins}\ \emph {et~al.}(2016)\citenamefont
  {Hopkins}, \citenamefont {Butler}, \citenamefont {Guttridge}, \citenamefont
  {Kemp}, \citenamefont {Freytag}, \citenamefont {Hinds}, \citenamefont
  {Tarbutt},\ and\ \citenamefont {Cornish}}]{Hopk2016}%
  \BibitemOpen
  \bibfield  {author} {\bibinfo {author} {\bibfnamefont {S.~A.}\ \bibnamefont
  {Hopkins}}, \bibinfo {author} {\bibfnamefont {K.}~\bibnamefont {Butler}},
  \bibinfo {author} {\bibfnamefont {A.}~\bibnamefont {Guttridge}}, \bibinfo
  {author} {\bibfnamefont {S.}~\bibnamefont {Kemp}}, \bibinfo {author}
  {\bibfnamefont {R.}~\bibnamefont {Freytag}}, \bibinfo {author} {\bibfnamefont
  {E.~A.}\ \bibnamefont {Hinds}}, \bibinfo {author} {\bibfnamefont {M.~R.}\
  \bibnamefont {Tarbutt}},\ and\ \bibinfo {author} {\bibfnamefont {S.~L.}\
  \bibnamefont {Cornish}},\ }\bibfield  {title} {\bibinfo {title} {A versatile
  dual-species zeeman slower for caesium and ytterbium},\ }\href@noop {}
  {\bibfield  {journal} {\bibinfo  {journal} {Review of Scientific
  Instruments}\ }\textbf {\bibinfo {volume} {87}},\ \bibinfo {pages} {043109}
  (\bibinfo {year} {2016})}\BibitemShut {NoStop}%
\bibitem [{\citenamefont {Molenaar}\ \emph {et~al.}(1997)\citenamefont
  {Molenaar}, \citenamefont {van~der Straten}, \citenamefont {Heideman},\ and\
  \citenamefont {Metcalf}}]{Mole1997}%
  \BibitemOpen
  \bibfield  {author} {\bibinfo {author} {\bibfnamefont {P.~A.}\ \bibnamefont
  {Molenaar}}, \bibinfo {author} {\bibfnamefont {P.}~\bibnamefont {van~der
  Straten}}, \bibinfo {author} {\bibfnamefont {H.~G.~M.}\ \bibnamefont
  {Heideman}},\ and\ \bibinfo {author} {\bibfnamefont {H.}~\bibnamefont
  {Metcalf}},\ }\bibfield  {title} {\bibinfo {title} {Diagnostic technique for
  zeeman-compensated atomic beam slowing: Technique and results},\ }\href@noop
  {} {\bibfield  {journal} {\bibinfo  {journal} {Phys. Rev. A}\ }\textbf
  {\bibinfo {volume} {55}},\ \bibinfo {pages} {605} (\bibinfo {year}
  {1997})}\BibitemShut {NoStop}%
\bibitem [{\citenamefont {Courtillot}\ \emph {et~al.}(2003)\citenamefont
  {Courtillot}, \citenamefont {Quessada}, \citenamefont {Kovacich},
  \citenamefont {Zondy}, \citenamefont {Landragin}, \citenamefont {Clairon},\
  and\ \citenamefont {Lemonde}}]{Court2003}%
  \BibitemOpen
  \bibfield  {author} {\bibinfo {author} {\bibfnamefont {I.}~\bibnamefont
  {Courtillot}}, \bibinfo {author} {\bibfnamefont {A.}~\bibnamefont
  {Quessada}}, \bibinfo {author} {\bibfnamefont {R.~P.}\ \bibnamefont
  {Kovacich}}, \bibinfo {author} {\bibfnamefont {J.-J.}\ \bibnamefont {Zondy}},
  \bibinfo {author} {\bibfnamefont {A.}~\bibnamefont {Landragin}}, \bibinfo
  {author} {\bibfnamefont {A.}~\bibnamefont {Clairon}},\ and\ \bibinfo {author}
  {\bibfnamefont {P.}~\bibnamefont {Lemonde}},\ }\bibfield  {title} {\bibinfo
  {title} {Efficient cooling and trapping of strontium atoms},\ }\href@noop {}
  {\bibfield  {journal} {\bibinfo  {journal} {Opt. Lett.}\ }\textbf {\bibinfo
  {volume} {28}},\ \bibinfo {pages} {468} (\bibinfo {year} {2003})}\BibitemShut
  {NoStop}%
\bibitem [{\citenamefont {Cheiney}\ \emph {et~al.}(2011)\citenamefont
  {Cheiney}, \citenamefont {Carraz}, \citenamefont {Bartoszek-Bober},
  \citenamefont {Faure}, \citenamefont {Vermersch}, \citenamefont {Fabre},
  \citenamefont {Gattobigio}, \citenamefont {Lahaye}, \citenamefont
  {Guéry-Odelin},\ and\ \citenamefont {Mathevet}}]{Chei2011}%
  \BibitemOpen
  \bibfield  {author} {\bibinfo {author} {\bibfnamefont {P.}~\bibnamefont
  {Cheiney}}, \bibinfo {author} {\bibfnamefont {O.}~\bibnamefont {Carraz}},
  \bibinfo {author} {\bibfnamefont {D.}~\bibnamefont {Bartoszek-Bober}},
  \bibinfo {author} {\bibfnamefont {S.}~\bibnamefont {Faure}}, \bibinfo
  {author} {\bibfnamefont {F.}~\bibnamefont {Vermersch}}, \bibinfo {author}
  {\bibfnamefont {C.~M.}\ \bibnamefont {Fabre}}, \bibinfo {author}
  {\bibfnamefont {G.~L.}\ \bibnamefont {Gattobigio}}, \bibinfo {author}
  {\bibfnamefont {T.}~\bibnamefont {Lahaye}}, \bibinfo {author} {\bibfnamefont
  {D.}~\bibnamefont {Guéry-Odelin}},\ and\ \bibinfo {author} {\bibfnamefont
  {R.}~\bibnamefont {Mathevet}},\ }\bibfield  {title} {\bibinfo {title} {A
  zeeman slower design with permanent magnets in a halbach configuration},\
  }\href@noop {} {\bibfield  {journal} {\bibinfo  {journal} {Review of
  Scientific Instruments}\ }\textbf {\bibinfo {volume} {82}},\ \bibinfo {pages}
  {063115} (\bibinfo {year} {2011})}\BibitemShut {NoStop}%
\bibitem [{\citenamefont {Qiang}\ \emph {et~al.}(2015)\citenamefont {Qiang},
  \citenamefont {Yi-Ge}, \citenamefont {Fang-Lin}, \citenamefont {Ye},
  \citenamefont {Bai-Ke}, \citenamefont {Fei}, \citenamefont {Er-Jun},
  \citenamefont {Tian-Chu},\ and\ \citenamefont {Zhan-Jun}}]{Qiang2015}%
  \BibitemOpen
  \bibfield  {author} {\bibinfo {author} {\bibfnamefont {W.}~\bibnamefont
  {Qiang}}, \bibinfo {author} {\bibfnamefont {L.}~\bibnamefont {Yi-Ge}},
  \bibinfo {author} {\bibfnamefont {G.}~\bibnamefont {Fang-Lin}}, \bibinfo
  {author} {\bibfnamefont {L.}~\bibnamefont {Ye}}, \bibinfo {author}
  {\bibfnamefont {L.}~\bibnamefont {Bai-Ke}}, \bibinfo {author} {\bibfnamefont
  {M.}~\bibnamefont {Fei}}, \bibinfo {author} {\bibfnamefont {Z.}~\bibnamefont
  {Er-Jun}}, \bibinfo {author} {\bibfnamefont {L.}~\bibnamefont {Tian-Chu}},\
  and\ \bibinfo {author} {\bibfnamefont {F.}~\bibnamefont {Zhan-Jun}},\
  }\bibfield  {title} {\bibinfo {title} {A longitudinal zeeman slower based on
  ring-shaped permanent magnets for a strontium optical lattice clock},\
  }\href@noop {} {\bibfield  {journal} {\bibinfo  {journal} {Chinese Physics
  Letters}\ }\textbf {\bibinfo {volume} {32}},\ \bibinfo {pages} {100701}
  (\bibinfo {year} {2015})}\BibitemShut {NoStop}%
\bibitem [{\citenamefont {Wodey}\ \emph {et~al.}(2021)\citenamefont {Wodey},
  \citenamefont {Rengelink}, \citenamefont {Meiners}, \citenamefont {Rasel},\
  and\ \citenamefont {Schlippert}}]{Wodey2021}%
  \BibitemOpen
  \bibfield  {author} {\bibinfo {author} {\bibfnamefont {E.}~\bibnamefont
  {Wodey}}, \bibinfo {author} {\bibfnamefont {R.~J.}\ \bibnamefont
  {Rengelink}}, \bibinfo {author} {\bibfnamefont {C.}~\bibnamefont {Meiners}},
  \bibinfo {author} {\bibfnamefont {E.~M.}\ \bibnamefont {Rasel}},\ and\
  \bibinfo {author} {\bibfnamefont {D.}~\bibnamefont {Schlippert}},\ }\bibfield
   {title} {\bibinfo {title} {A robust, high-flux source of laser-cooled
  ytterbium atoms},\ }\href@noop {} {\bibfield  {journal} {\bibinfo  {journal}
  {Journal of Physics B: Atomic, Molecular and Optical Physics}\ }\textbf
  {\bibinfo {volume} {54}},\ \bibinfo {pages} {035301} (\bibinfo {year}
  {2021})}\BibitemShut {NoStop}%
\bibitem [{\citenamefont {Metcalf}\ and\ \citenamefont {van~der
  Straten}(2001)}]{Metcalf2001}%
  \BibitemOpen
  \bibfield  {author} {\bibinfo {author} {\bibfnamefont {H.}~\bibnamefont
  {Metcalf}}\ and\ \bibinfo {author} {\bibfnamefont {P.}~\bibnamefont {van~der
  Straten}},\ }\href@noop {} {\emph {\bibinfo {title} {Laser Cooling and
  Trapping}}},\ Graduate Texts in Contemporary Physics\ (\bibinfo  {publisher}
  {Springer New York},\ \bibinfo {year} {2001})\BibitemShut {NoStop}%
\bibitem [{\citenamefont {Ramsey}(1956)}]{ramsey1956}%
  \BibitemOpen
  \bibfield  {author} {\bibinfo {author} {\bibfnamefont {N.}~\bibnamefont
  {Ramsey}},\ }\href@noop {} {\emph {\bibinfo {title} {Molecular Beams}}}\
  (\bibinfo  {publisher} {Clarendon Press},\ \bibinfo {year}
  {1956})\BibitemShut {NoStop}%
\bibitem [{\citenamefont {Heubner}\ and\ \citenamefont
  {Markiewicz}(2000)}]{huebner2000}%
  \BibitemOpen
  \bibfield  {author} {\bibinfo {author} {\bibfnamefont {W.}~\bibnamefont
  {Heubner}}\ and\ \bibinfo {author} {\bibfnamefont {W.}~\bibnamefont
  {Markiewicz}},\ }\bibfield  {title} {\bibinfo {title} {The temperature and
  bulk flow speed of a gas effusing or evaporating from a surface into a void
  after reestablishment of collisional equilibrium},\ }\href@noop {} {\bibfield
   {journal} {\bibinfo  {journal} {Icarus}\ }\textbf {\bibinfo {volume}
  {148}},\ \bibinfo {pages} {594} (\bibinfo {year} {2000})}\BibitemShut
  {NoStop}%
\bibitem [{Note1()}]{Note1}%
  \BibitemOpen
  \bibinfo {note} {A transverse field slower requires the slowing laser to be
  linearly polarized perpendicular to the magnetic field; therefore, only one
  helicity of the resulting $\sigma ^{+}$ and $\sigma ^{-}$ polarization
  superposition is useful for slowing \cite
  {Mele2004,Ovch2007,Hill2014}}\BibitemShut {NoStop}%
\bibitem [{\citenamefont {Bagnato}\ \emph {et~al.}(1991)\citenamefont
  {Bagnato}, \citenamefont {Salomon}, \citenamefont {Marega},\ and\
  \citenamefont {Zilio}}]{Bagn1991}%
  \BibitemOpen
  \bibfield  {author} {\bibinfo {author} {\bibfnamefont {V.~S.}\ \bibnamefont
  {Bagnato}}, \bibinfo {author} {\bibfnamefont {C.}~\bibnamefont {Salomon}},
  \bibinfo {author} {\bibfnamefont {E.}~\bibnamefont {Marega}},\ and\ \bibinfo
  {author} {\bibfnamefont {S.~C.}\ \bibnamefont {Zilio}},\ }\bibfield  {title}
  {\bibinfo {title} {Influence of adiabatic following and optical pumping in
  the production of an intense steady flux of slow atoms},\ }\href@noop {}
  {\bibfield  {journal} {\bibinfo  {journal} {J. Opt. Soc. Am. B}\ }\textbf
  {\bibinfo {volume} {8}},\ \bibinfo {pages} {497} (\bibinfo {year}
  {1991})}\BibitemShut {NoStop}%
\bibitem [{\citenamefont {Joffe}\ \emph {et~al.}(1993)\citenamefont {Joffe},
  \citenamefont {Ketterle}, \citenamefont {Martin},\ and\ \citenamefont
  {Pritchard}}]{Joff1993}%
  \BibitemOpen
  \bibfield  {author} {\bibinfo {author} {\bibfnamefont {M.~A.}\ \bibnamefont
  {Joffe}}, \bibinfo {author} {\bibfnamefont {W.}~\bibnamefont {Ketterle}},
  \bibinfo {author} {\bibfnamefont {A.}~\bibnamefont {Martin}},\ and\ \bibinfo
  {author} {\bibfnamefont {D.~E.}\ \bibnamefont {Pritchard}},\ }\bibfield
  {title} {\bibinfo {title} {Transverse cooling and deflection of an atomic
  beam inside a zeeman slower},\ }\href@noop {} {\bibfield  {journal} {\bibinfo
   {journal} {J. Opt. Soc. Am. B}\ }\textbf {\bibinfo {volume} {10}},\ \bibinfo
  {pages} {2257} (\bibinfo {year} {1993})}\BibitemShut {NoStop}%
\bibitem [{\citenamefont {McClelland}\ and\ \citenamefont
  {Hanssen}(2006)}]{McCl2006}%
  \BibitemOpen
  \bibfield  {author} {\bibinfo {author} {\bibfnamefont {J.~J.}\ \bibnamefont
  {McClelland}}\ and\ \bibinfo {author} {\bibfnamefont {J.~L.}\ \bibnamefont
  {Hanssen}},\ }\bibfield  {title} {\bibinfo {title} {Laser cooling without
  repumping: A magneto-optical trap for erbium atoms},\ }\href@noop {}
  {\bibfield  {journal} {\bibinfo  {journal} {Phys. Rev. Lett.}\ }\textbf
  {\bibinfo {volume} {96}},\ \bibinfo {pages} {143005} (\bibinfo {year}
  {2006})}\BibitemShut {NoStop}%
\bibitem [{\citenamefont {Marti}\ \emph {et~al.}(2010)\citenamefont {Marti},
  \citenamefont {Olf}, \citenamefont {Vogt}, \citenamefont {\"Ottl},\ and\
  \citenamefont {Stamper-Kurn}}]{Marti2010}%
  \BibitemOpen
  \bibfield  {author} {\bibinfo {author} {\bibfnamefont {G.~E.}\ \bibnamefont
  {Marti}}, \bibinfo {author} {\bibfnamefont {R.}~\bibnamefont {Olf}}, \bibinfo
  {author} {\bibfnamefont {E.}~\bibnamefont {Vogt}}, \bibinfo {author}
  {\bibfnamefont {A.}~\bibnamefont {\"Ottl}},\ and\ \bibinfo {author}
  {\bibfnamefont {D.~M.}\ \bibnamefont {Stamper-Kurn}},\ }\bibfield  {title}
  {\bibinfo {title} {Two-element zeeman slower for rubidium and lithium},\
  }\href@noop {} {\bibfield  {journal} {\bibinfo  {journal} {Phys. Rev. A}\
  }\textbf {\bibinfo {volume} {81}},\ \bibinfo {pages} {043424} (\bibinfo
  {year} {2010})}\BibitemShut {NoStop}%
\bibitem [{\citenamefont {Ben~Ali}(2016)}]{Ali2016}%
  \BibitemOpen
  \bibfield  {author} {\bibinfo {author} {\bibfnamefont {D.}~\bibnamefont
  {Ben~Ali}},\ }\emph {\bibinfo {title} {{Conception et construction d'une
  exp{\'e}rience d'atomes froids : vers un condensat de sodium sur puce}}},\
  \href@noop {} {Ph.D. thesis},\ \bibinfo  {school} {{Universit{\'e} Paris 13,
  Sorbonne Paris Cit{\'e}}} (\bibinfo {year} {2016})\BibitemShut {NoStop}%
\bibitem [{\citenamefont {G{\"u}nter}()}]{gunter2004}%
  \BibitemOpen
  \bibfield  {author} {\bibinfo {author} {\bibfnamefont {K.~J.}\ \bibnamefont
  {G{\"u}nter}},\ }\href@noop {} {\bibinfo {title} {Design and implementation
  of a zeeman slower for {$^{87}\mathrm{Rb}$}}},\ \bibinfo {note} {available
  online at http://www.kenneth.ch/atoms/slower.pdf (2004)}\BibitemShut
  {NoStop}%
\bibitem [{\citenamefont {Bell}\ \emph {et~al.}(2010)\citenamefont {Bell},
  \citenamefont {Junker}, \citenamefont {Jasperse}, \citenamefont {Turner},
  \citenamefont {Lin}, \citenamefont {Spielman},\ and\ \citenamefont
  {Scholten}}]{Bell2010}%
  \BibitemOpen
  \bibfield  {author} {\bibinfo {author} {\bibfnamefont {S.~C.}\ \bibnamefont
  {Bell}}, \bibinfo {author} {\bibfnamefont {M.}~\bibnamefont {Junker}},
  \bibinfo {author} {\bibfnamefont {M.}~\bibnamefont {Jasperse}}, \bibinfo
  {author} {\bibfnamefont {L.~D.}\ \bibnamefont {Turner}}, \bibinfo {author}
  {\bibfnamefont {Y.-J.}\ \bibnamefont {Lin}}, \bibinfo {author} {\bibfnamefont
  {I.~B.}\ \bibnamefont {Spielman}},\ and\ \bibinfo {author} {\bibfnamefont
  {R.~E.}\ \bibnamefont {Scholten}},\ }\bibfield  {title} {\bibinfo {title} {A
  slow atom source using a collimated effusive oven and a single-layer variable
  pitch coil zeeman slower},\ }\href@noop {} {\bibfield  {journal} {\bibinfo
  {journal} {Review of Scientific Instruments}\ }\textbf {\bibinfo {volume}
  {81}},\ \bibinfo {pages} {013105} (\bibinfo {year} {2010})}\BibitemShut
  {NoStop}%
\bibitem [{\citenamefont {Witte}\ \emph {et~al.}(1992)\citenamefont {Witte},
  \citenamefont {Kisters}, \citenamefont {Riehle},\ and\ \citenamefont
  {Helmcke}}]{Witte1992}%
  \BibitemOpen
  \bibfield  {author} {\bibinfo {author} {\bibfnamefont {A.}~\bibnamefont
  {Witte}}, \bibinfo {author} {\bibfnamefont {T.}~\bibnamefont {Kisters}},
  \bibinfo {author} {\bibfnamefont {F.}~\bibnamefont {Riehle}},\ and\ \bibinfo
  {author} {\bibfnamefont {J.}~\bibnamefont {Helmcke}},\ }\bibfield  {title}
  {\bibinfo {title} {Laser cooling and deflection of a calcium atomic beam},\
  }\href@noop {} {\bibfield  {journal} {\bibinfo  {journal} {J. Opt. Soc. Am.
  B}\ }\textbf {\bibinfo {volume} {9}},\ \bibinfo {pages} {1030} (\bibinfo
  {year} {1992})}\BibitemShut {NoStop}%
\bibitem [{\citenamefont {Lu}\ \emph {et~al.}(2010)\citenamefont {Lu},
  \citenamefont {Youn},\ and\ \citenamefont {Lev}}]{Lu2010}%
  \BibitemOpen
  \bibfield  {author} {\bibinfo {author} {\bibfnamefont {M.}~\bibnamefont
  {Lu}}, \bibinfo {author} {\bibfnamefont {S.~H.}\ \bibnamefont {Youn}},\ and\
  \bibinfo {author} {\bibfnamefont {B.~L.}\ \bibnamefont {Lev}},\ }\bibfield
  {title} {\bibinfo {title} {Trapping ultracold dysprosium: A highly magnetic
  gas for dipolar physics},\ }\href@noop {} {\bibfield  {journal} {\bibinfo
  {journal} {Phys. Rev. Lett.}\ }\textbf {\bibinfo {volume} {104}},\ \bibinfo
  {pages} {063001} (\bibinfo {year} {2010})}\BibitemShut {NoStop}%
\bibitem [{Note2()}]{Note2}%
  \BibitemOpen
  \bibinfo {note} {A special case of an efficient transverse field spin-flip
  slower was demonstrated with Sr \cite {Hill2014}, which has a $J=0$ lower
  cooling state and therefore does not suffer the depumping issues discussed in
  Ref. \cite {Ali2016}.}\BibitemShut {Stop}%
\bibitem [{\citenamefont {Barrett}\ \emph {et~al.}(1991)\citenamefont
  {Barrett}, \citenamefont {Dapore-Schwartz}, \citenamefont {Ray},\ and\
  \citenamefont {Lafyatis}}]{Barr1991}%
  \BibitemOpen
  \bibfield  {author} {\bibinfo {author} {\bibfnamefont {T.~E.}\ \bibnamefont
  {Barrett}}, \bibinfo {author} {\bibfnamefont {S.~W.}\ \bibnamefont
  {Dapore-Schwartz}}, \bibinfo {author} {\bibfnamefont {M.~D.}\ \bibnamefont
  {Ray}},\ and\ \bibinfo {author} {\bibfnamefont {G.~P.}\ \bibnamefont
  {Lafyatis}},\ }\bibfield  {title} {\bibinfo {title} {Slowing atoms with
  ${\mathrm{\ensuremath{\sigma}}}^{\mathrm{\ensuremath{-}}}$ polarized light},\
  }\href@noop {} {\bibfield  {journal} {\bibinfo  {journal} {Phys. Rev. Lett.}\
  }\textbf {\bibinfo {volume} {67}},\ \bibinfo {pages} {3483} (\bibinfo {year}
  {1991})}\BibitemShut {NoStop}%
\bibitem [{Note3()}]{Note3}%
  \BibitemOpen
  \bibinfo {note} {In the transverse direction, we observe a linewidth of 29
  MHz, which is slightly broader than the 20.7 MHz natural
  linewidth.}\BibitemShut {Stop}%
\bibitem [{\citenamefont {Stancik}\ and\ \citenamefont
  {Brauns}(2008)}]{Stan2008}%
  \BibitemOpen
  \bibfield  {author} {\bibinfo {author} {\bibfnamefont {A.~L.}\ \bibnamefont
  {Stancik}}\ and\ \bibinfo {author} {\bibfnamefont {E.~B.}\ \bibnamefont
  {Brauns}},\ }\bibfield  {title} {\bibinfo {title} {A simple asymmetric
  lineshape for fitting infrared absorption spectra},\ }\href@noop {}
  {\bibfield  {journal} {\bibinfo  {journal} {Vibrational Spectroscopy}\
  }\textbf {\bibinfo {volume} {47}},\ \bibinfo {pages} {66} (\bibinfo {year}
  {2008})}\BibitemShut {NoStop}%
\bibitem [{Note4()}]{Note4}%
  \BibitemOpen
  \bibinfo {note} {A characterization of the 410~nm state preparation lasers is
  not described here because they do not affect our Zeeman slower efficiency
  (only the overall atom number)}\BibitemShut {NoStop}%
\bibitem [{\citenamefont {Petzold}\ \emph {et~al.}(2018)\citenamefont
  {Petzold}, \citenamefont {Kaebert}, \citenamefont {Gersema}, \citenamefont
  {Poll}, \citenamefont {Reinhardt}, \citenamefont {Siercke},\ and\
  \citenamefont {Ospelkaus}}]{Petz2018}%
  \BibitemOpen
  \bibfield  {author} {\bibinfo {author} {\bibfnamefont {M.}~\bibnamefont
  {Petzold}}, \bibinfo {author} {\bibfnamefont {P.}~\bibnamefont {Kaebert}},
  \bibinfo {author} {\bibfnamefont {P.}~\bibnamefont {Gersema}}, \bibinfo
  {author} {\bibfnamefont {T.}~\bibnamefont {Poll}}, \bibinfo {author}
  {\bibfnamefont {N.}~\bibnamefont {Reinhardt}}, \bibinfo {author}
  {\bibfnamefont {M.}~\bibnamefont {Siercke}},\ and\ \bibinfo {author}
  {\bibfnamefont {S.}~\bibnamefont {Ospelkaus}},\ }\bibfield  {title} {\bibinfo
  {title} {Type-ii zeeman slowing: Characterization and comparison to
  conventional radiative beam-slowing schemes},\ }\href@noop {} {\bibfield
  {journal} {\bibinfo  {journal} {Phys. Rev. A}\ }\textbf {\bibinfo {volume}
  {98}},\ \bibinfo {pages} {063408} (\bibinfo {year} {2018})}\BibitemShut
  {NoStop}%
\bibitem [{\citenamefont {Steck}()}]{steck2021}%
  \BibitemOpen
  \bibfield  {author} {\bibinfo {author} {\bibfnamefont {D.}~\bibnamefont
  {Steck}},\ }\href@noop {} {\bibinfo {title} {Quantum and atom optics}},\
  \bibinfo {note} {available online at http://steck.us/teaching (revision
  0.13.10, 22 September 2021)}\BibitemShut {NoStop}%
\end{thebibliography}%

\end{document}